\documentclass{article}

\usepackage{geometry}
\usepackage[utf8]{inputenc} 
\usepackage[T1]{fontenc}    
\usepackage{hyperref}       
\usepackage{url}            
\usepackage{booktabs}       
\usepackage{amsfonts}       
\usepackage{nicefrac}       
\usepackage{microtype}      
\usepackage{xcolor}         
\usepackage{graphicx}
\usepackage{subcaption}
\usepackage{mdframed}
\usepackage{makecell}

\usepackage{tikz}
\usepackage{pgfplots}

\usepackage{amsmath}       
\usepackage{amsthm}
\theoremstyle{definition}
\newtheorem{theorem}{Theorem}[section]
\newtheorem{lemma}{Lemma}[section]

\newtheorem{problem}{Problem}[section]
\newtheorem{corollary}{Corollary}[section]

\usepackage[ruled,linesnumbered]{algorithm2e}
\usepackage{wrapfig}

\newcommand{\warning}[1]{{\color{black}{#1}}}

\title{Data Debugging is NP-hard for Classifiers Trained with SGD}

\author{
  Zizheng Guo\\
  Harbin Institute of Technology\\
  \texttt{zguo.research@gmail.com}
  \and
  Pengyu Chen\\
  Harbin Institute of Technology\\
  \texttt{pchen.research@gmail.com}
  \and
  Yanzhang Fu\\
  Harbin Institute of Technology\\
  \texttt{23S003008@stu.hit.edu.cn}
  \and
  Dongjing Miao\\
  Harbin Institute of Technology\\
  \texttt{miaodongjing@hit.edu.cn}\\
}

\usepackage{lipsum}
\pgfplotsset{compat=1.18}
\begin{document}

\maketitle

\begin{abstract}

Data debugging is to find a subset of the training data such that the model obtained by retraining on the subset has a better accuracy.
A bunch of heuristic approaches are proposed, however, none of them are guaranteed to solve this problem effectively.
This leaves an open issue whether there exists an efficient algorithm to find the subset such that the model obtained by retraining on it has a better accuracy.
To answer this open question and provide theoretical basis for further study on developing better algorithms for data debugging, we investigate the computational complexity of the problem named \textsc{Debuggable}.
Given a machine learning model $\mathcal{M}$ obtained by training on dataset $D$ and a test instance $(\mathbf{x}_\text{test},y_\text{test})$ where $\mathcal{M}(\mathbf{x}_\text{test})\neq y_\text{test}$, \textsc{Debuggable} is to determine whether there exists a subset $D^\prime$ of $D$ such that the model $\mathcal{M}^\prime$ obtained by retraining on $D^\prime$ satisfies $\mathcal{M}^\prime(\mathbf{x}_\text{test})=y_\text{test}$.
To cover a wide range of commonly used models, we take SGD-trained linear classifier as the model and derive the following main results.
(1) If the loss function and the dimension of the model are not fixed, \textsc{Debuggable} is NP-complete regardless of the training order in which all the training samples are processed during SGD.
(2) For hinge-like loss functions, a comprehensive analysis on the computational complexity of \textsc{Debuggable} is provided;
(3) If the loss function is a linear function, \textsc{Debuggable} can be solved in linear time, that is, data debugging can be solved easily in this case.
These results not only highlight the limitations of current approaches but also offer new insights into data debugging.
\end{abstract}

\section{Introduction}

Given a machine learning model, data debugging is to find a subset of the training data such that the model will have a better accuracy if retrained on that subset \cite{nips19-data-cleansing-with-sgd}.
Data debugging serves as a popular method of both data cleaning and machine learning interpretation.
In the context of data cleaning, data debugging (\textit{a.k.a.} training data debugging \cite{Query2.0} or data cleansing \cite{nips19-data-cleansing-with-sgd}) can be used to improve the quality of the training data by removing the flaws leading to mispredictions \cite{karlaš2022data-shapley,Neutatz2021FromCB, CleanML}. 
When it comes to ML interpretation, data debugging locates the part of the training data responsible for unexpected predictions of an ML model. Therefore it is also studied as a training data-based (\textit{a.k.a.} instance-based \cite{influence-func-answer}) interpretation, which is crucial for helping system developers and ML practitioners to debug ML system by reporting the harmful part of training data \cite{SIGMOD22-interp-data-based-exp}.





To solve the data debugging problem, existing researches adopt a two-phase score-based heuristic approach \cite{Query2.0}.
In the first phase, a score representing the estimated impact on the model accuracy is assigned to each training sample in the training data. It is hoped that the harmful part of training data gets a lower score than the other part. In the second phase, training samples with lower scores are removed greedily and the model is retrained on the modified training data.
The two phases are carried out iteratively until a well-trained model is obtained.
Most of the related works focus on developing algorithms to estimate the scores efficiently in the first phase \cite{ICML-influence,Khanna2018InterpretingBB,nips19-influence-liang,SO-IF,guo-etal-2021-fastif,Ghorbani2019DataSE,Jia2019TowardsED,sca-util-shap,shap-knn-proxy}, but rarely study the effectiveness of the entire two-phase approach.

Since it is computationally intractable to estimate the score for all possible subsets of the training data, it is often assumed that the score representing the impact of a subset is approximately equal to the sum of the scores of each individual training samples from the subset. However, Koh et. al. \cite{nips19-influence-liang} showed this is not always the case. For a bunch of subsets sampled from the training data, they empirically studied the difference between the estimated impact and the actual impact of each subset by taking influence functions as the scoring method. The estimated impact is calculated by summing up the score by influence function of each training samples in the subset, and the actual impact is measured by the improvement of accuracy of the model retrained after removing the subset from training data. They found that the estimated impact tends to underestimate the actual impact. Removing a large number of training samples could result in a large deviation between estimated and actual impacts.
Although an upper bound of the deviation under certain assumptions has been derived, it is still unknown whether the deviation can be reduced or eliminated efficiently.

The above deviation also poses challenges to the effectiveness of the entire approach. Suppose the influence function is adopted as the scoring method, the accuracy of the model is not guaranteed to improve due to the deviation reported in  \cite{nips19-influence-liang} if a large group of training samples are removed during each iteration. 
Moreover, there is no theoretical analysis for the effectiveness of the greedy approach in the second phase. Even if only one training sample is removed during each iteration of the two-phase approach, the accuracy of the model is still not guaranteed to be improved. The effectiveness of the entire two-phase approach is therefore not assured. This leaves the following open problem:
 \begin{problem}\label{open-problem}
     Is there an efficient algorithm to find the subset of the training data, such that the model obtained by retraining on it has a better accuracy?
 \end{problem}

The computational complexity results presented in this paper demonstrate that it is unlikely to solve the data debugging problem efficiently in polynomial time.
To figure out its hardness, we study the problem  \textsc{Debuggable} which is the decision version of data debugging when the test set consists of only one instance.
Formally, \textsc{Debuggable} is defined as follows: 

\begin{problem}[\textsc{Debuggable}]
    Given a classifier $\mathcal{M}$, its training data $T$, a test instance $(\mathbf{x},y)$. Is there a $T^\prime\subseteq T$, such that $\mathcal{M}$ predicts $y$ on $\mathbf{x}$ if retrained on $T^\prime$?
\end{problem}
Basically, we prove that \textsc{Debuggable} is NP-complete, which means data debugging is unlikely to be solved in polynomial time.
This result answers the open question mentioned above directly, this is, the large deviation of estimated impacts \cite{nips19-influence-liang} cannot be reduced or eliminated efficiently.
This is because if the impact of a subset of the training data could be accurately estimated as the sum of the impact of each training sample in the subset, data debugging can be solved in polynomial time, which is impossible unless P=NP.


Although \textsc{Debuggable} is generally intractable, we still hope to develop efficient algorithms tailored to specific cases. Thus it is necessary to figure out the root cause of the hardness for \textsc{Debuggable}. 
Previous research are always conducted based on the belief that the complexity of data debugging is due to the chosen model architecture is complicated.
However, we show that at least for models trained by stochastic gradient descent (SGD), the hardness stems from the hyper-parameter configuration selected for the SGD training, which was not yet aware of by previous work.
To cover a wide range of commonly used machine learning models, we take linear classifiers as the model and show that even for linear classifiers, \textsc{Debuggable} is NP-hard as long as they are trained by SGD.  Moreover, we provided a comprehensive analysis on hyper-parameter configurations that affect the computational complexity of \textsc{Debuggable}, including the loss function, the model dimension and the training order. Training order, \emph{a.k.a.} training data order \cite{training-data-order} or order of training samples \cite{order-of-training-sample}, refers to the order in which each training sample is considered during the SGD.
Detailed complexity results are shown in Table \ref{tab:complexity-results}.

Our contribution can be concluded as follows:
\begin{itemize}
    \item We studied the computational complexity of data debugging and showed that data debugging is NP-hard for linear classifiers in the general setting for \emph{all possible training orders}.

    \item We studied the complexity of \textsc{Debuggable} when the loss is fixed as the hinge-like function. For 2 or higher dimension, \textsc{Debuggable} is NP-complete when the training order is adversarially chosen; For one-dimensional cases, \textsc{Debuggable} can be NP-hard when the interception $\beta<0$, and is solvable in linear time when $\beta \ge 0$.

    \item  We proved that \textsc{Debuggable} is solvable in linear time when the loss function is linear.

\end{itemize}

Moreover, we have a discussion on the implications of these complexity results for machine learning interpretability and data quality, as well as limitations of score-based greedy methods. Our results suggest the further study as follows. (1) It is better to characterize the training sample and find the criterion which can be used to decide the existence of efficient algorithms; (2) Designing algorithms with CSP-solver is a potential way to solve data debugging more efficiently than the brute-force one; (3) Developing random algorithms is a potential way to solve data debugging successfully with high probability. 

    \begin{table}
    \centering
    \caption{Computational complexity of the data debugging problem}\label{tab:complexity-results}
        \begin{tabular}{cccc}
        \toprule
        Loss Function & Dimension & Training Order & Complexity\\
        \midrule
        Not Fixed & Not Fixed & - & NP-hard\\\
        Hinge-like  & $\ge 2$ & Adversarially Chosen& NP-hard \\
        Hinge-like, $\beta< 0$ & $1$ & Adversarially Chosen&NP-hard\\
        Hinge-like, $\beta\ge 0$ & $1$ & - & Linear Time\\
        Linear & - & -& Linear Time\\
        \bottomrule
        \end{tabular}
    \end{table}

\subsection{Related Works}
The solution of data debugging has applications in database query results reliability enhancement \cite{Query2.0,FROG}, training data cleaning \cite{nips19-data-cleansing-with-sgd} and machine learning interpretation\cite{Khanna2018InterpretingBB, ICML-influence, nips19-influence-liang, brunet2019understanding, wang2019repairing}. Existing works on data debugging mainly adopt a two-phase approach, which scores the training samples in the first phase and greedily deletes training samples with lower scores in the second phase. 
Most of the research focus on the first phase.
There are mainly two ways of scoring adopted for data debugging in practice.
Leave-one-out (LOO) retraining is a widely studied way, which evaluates the contribution of a training sample through the difference in the model's accuracy trained without that training sample.
To avoid the cost of model retraining, Koh and Liang took influence functions as an approximation of LOO \cite{ICML-influence}. After that, various extensions and improvements of the influence function based method are proposed, such as Fisher kernel \cite{Khanna2018InterpretingBB}, influence function for group impacts \cite{nips19-influence-liang}, second-order approximations \cite{SO-IF} and scalable influence functions \cite{guo-etal-2021-fastif}.
Another way is Shapley-based scoring, where the impact of a training sample is measured by its average marginal contribution to all subsets of the training data \cite{Ghorbani2019DataSE}. 
Since Shapley-base scoring suffers from expensive computational cost \cite{Deng1994OnTC}, recent works focus on techniques that efficiently estimate the Shapley value, including Monte-Carlo sampling \cite{Ghorbani2019DataSE}, group testing \cite{Jia2019TowardsED,sca-util-shap} and using proxy models such as $k$-NN  \cite{shap-knn-proxy,karlaš2022data-shapley}. However, those methods do not admit any theoretical guarantee on the effectiveness. This paper discusses the limitations of the above methods and suggests some future directions on data debugging.


\section{Preliminaries and Problem Definition}
\textbf{Linear classifiers.} Formally, a (binary) linear classifier is a function $\lambda_\mathbf{w}:\mathbb{R}^d\rightarrow\{-1,1\}$, where $d$ is called its \emph{dimension} and $\mathbf{w}\in \mathbb{R}^{d}$ its parameter. Without loss of generality, the bias term of a linear classifier is set as zero in this paper. All vectors in this paper are assumed to be \emph{column} vectors. For an input $\mathbf{x}$, the value of $\lambda_\mathbf{w}$ is defined as
\begin{equation*}
\lambda_\mathbf{w}(\mathbf{x})=\begin{cases}
    1 ~&\text{if} ~ \mathbf{w}^\top \mathbf{x} \ge 0\\
   -1 ~&\text{otherwise.}
\end{cases}
\end{equation*}
We denote the class of linear models as $\Lambda$.

\textbf{Training data.} 
A \textit{training sample} is a pair $(\mathbf{x},y)$ in which $\mathbf{x}\in \mathbb{R}^d$ is the input and $y\in \{-1,1\}$ is the label of $\mathbf{x}$. 
The \textit{training data} is a multiset of training samples.
\warning{We employ $\textbf{w}\xrightarrow{T}\textbf{w}'$ to denote that the parameter $\textbf{w}'$ is obtained by training the parameter $\textbf{w}$ on the training data $T$, and employ $\textbf{w}\xrightarrow{(\textbf{x},y)}\textbf{w}'$ to denote that $\textbf{w}'$ is obtained by training $\textbf{w}$ on the training sample $(\textbf{x},y)$.}

\textbf{Loss functions and learning rates.} 
Binary linear classifiers typically use unary functions on $y\mathbf{w}^\top\mathbf{x}$ as their loss functions \cite{loss-survey}.
Therefore we only consider loss functions of the form $\mathcal{L}: y\mathbf{w}^\top\mathbf{x} \mapsto \mathbb{R}$ for the rest of the paper. 

The \textit{linear} loss is in the form of 
\begin{equation*}
    \mathcal{L}_\texttt{lin}(y\mathbf{w}^\top\mathbf{x})=-\alpha (y\mathbf{w}^\top \mathbf{x} + \beta).
\end{equation*}

The \textit{hinge-like} loss function is defined as the following form
\begin{equation*}
    \mathcal{L}_\texttt{hinge}(y\mathbf{w}^\top\mathbf{x}) = \begin{cases}
        -\alpha (y\mathbf{w}^\top\mathbf{x} - \beta), & y\mathbf{w}^\top\mathbf{x}<\beta\\
        0, & \text{otherwise.}
    \end{cases}
\end{equation*}
We call $\beta$ as the \emph{interception} of $\mathcal{L}_\texttt{hinge}$.
We represent the learning rate of a model using a vector $\boldsymbol{\eta}=(\eta_1,\dots,\eta_{d})$, where $\eta_i \ge 0$ and each parameter $w_i$ can be updated with the corresponding learning rate $\eta_i$.

\textbf{Stochastic gradient descent.}
The stochastic gradient descent (SGD) method updates parameter $\mathbf{w}$ from its initial value $\mathbf{w}^{(0)}$ through several epochs.
During each epoch, the SGD goes through the entire set of training samples in some training order through several iterations. 
The training order is defined as a sequence of training samples, in the form of $(\mathbf{x}_1,y_1)\dots (\mathbf{x}_n,y_n)$. For $1\le i<j\le n$, $(\mathbf{x}_i,y_i)$ is considered before $(\mathbf{x}_j,y_j)$ during the SGD.
We use $w_i$ to denote the $i$-th coordinate of $\mathbf{w}$.
We also use $\mathbf{w}^{(e,k)}$ to denote the value of $\mathbf{w}$ at the end of $k$-th iteration of epoch $e$ and use $\mathbf{w}^{(e)}$ to denote the value of $\mathbf{w}$ after the end of epoch $e$.
Assuming $(\mathbf{x},y)$ to be the training sample considered at iteration $k$, the stochastic gradient descent (SGD) method updates parameter $w_i$ for each $i$ by
\begin{equation}\label{sgd-update}
    w_i^{(e,k)} \leftarrow w_i^{(e,k-1)} - \eta_i \cdot \frac{\partial \mathcal{L}(y(\mathbf{w}^{(e,k-1)})^\top\mathbf{x})}{\partial w_i}
\end{equation}
In other words, we have
\begin{equation*}
    \mathbf{w}^{(e,k)}\leftarrow \mathbf{w}^{(e,k-1)}-\boldsymbol{\eta}\otimes \nabla \mathcal{L}(y(\mathbf{w}^{(e,k-1)})^\top\mathbf{x})
\end{equation*}
 where $\boldsymbol{\eta} \otimes  \nabla\mathcal{L}=(\eta_1 \frac{\partial \mathcal{L}}{\partial w_1},\dots,\eta_d \frac{\partial \mathcal{L}}{\partial w_d})$ is the Hadamard product.
We say a training sample $\mathbf{x}$ is \emph{activated} at iteration $k$ during epoch $e$ if $\nabla\mathcal{L}(y(\mathbf{w}^{(e,k-1)})^\top\mathbf{x})\neq 0$.
The SGD terminates at the end of epoch $e$ if $\| \mathbf{w}^{(e-1)}-\mathbf{w}^{(e)}\|< \varepsilon$ for threshold $\varepsilon$ or $e$ reached some predetermined value. We denote $\mathbf{w}^*=\mathbf{w}^{(e)}$.
A linear classifier trained by SGD with the meta-parameters mentioned above is denoted as
$\texttt{SGD}_\Lambda(\mathcal{L},\boldsymbol{\eta},\varepsilon,T)=\lambda_{\mathbf{w}^*}$. 
With a slight abuse of notation, we define $\texttt{SGD}_\Lambda(\mathcal{L},\boldsymbol{\eta},\varepsilon,T,\mathbf{x})=\lambda_{\mathbf{w}^*}(\mathbf{x})$. We also use $\texttt{SGD}_\Lambda(T,\mathbf{x})$ to avoid cluttering when the context is clear.

\textbf{Problem definition.} With the above definitions, \textsc{Debuggable} for SGD-trained linear classifiers can be formalized as follows:

    \begin{center}
    \begin{minipage}{0.8\textwidth}
    \begin{mdframed}[
        linewidth=1pt,
        roundcorner=5pt
    ]
    \textsc{Debuggable-Lin}
    
    \textbf{Input:} Training data $T$, loss function 
    $\mathcal{L}$, initial parameter $\mathbf{w}^{(0)}$, learning rate $\boldsymbol{\eta}$, threshold $\varepsilon$ and instance $(\mathbf{x}_\text{test},y_\text{test})$.

    \textbf{Output:}
        ``Yes'': if $\exists\Delta\subseteq T$ such that $\texttt{SGD}_\Lambda(\mathcal{L},\boldsymbol{\eta},\varepsilon,T\setminus\Delta,\mathbf{x}_\text{test})=y_\text{test}$;\\
    \hspace*{3.8em}``No'': otherwise.
    \end{mdframed}
    \end{minipage}
    \end{center}    
   
    We say $\texttt{SGD}_\Lambda(\mathcal{L},\boldsymbol{\eta},\varepsilon,T)$
    is \emph{debuggable} on $(\mathbf{x}_\text{test},y_\text{test})$ if $(\mathcal{L}, \mathbf{w}^{(0)},\boldsymbol{\eta},\varepsilon,T,\mathbf{x}_\text{test},y_\text{test})$
    is a yes-instance of \textsc{Debuggable-Lin}, and not \emph{debuggable} on $(\mathbf{x}_\text{test},y_\text{test})$ otherwise.

\section{Results for Unfixed Loss Functions}\label{gd-hardness}
    In this section, we prove the NP-hardness of \textsc{Debuggable-Lin}. Intuitively, \textsc{Debuggable-Lin} is to determine whether there exists a subset $T'\subseteq T$ where activated training samples within $T'$ drive the parameter $\mathbf{w}$ toward the region defined by $y_\text{test}\mathbf{w}^\top\mathbf{x}_\text{test}>0$. 
    The activation of training samples depends on the complex interaction between the training data and the model.
    
    \begin{theorem}\label{GD-NPC}
        \textsc{Debuggable-Lin} is NP-hard for all training orders.
    \end{theorem}
    
    We only show the proof sketch and leave the details in the appendix. 

    \begin{proof}[Proof Sketch]
         We build a reduction from an NP-hard problem \textsc{Monotone 1-in-3 SAT} \cite{Computability-book}:
   
    \begin{center}
    \begin{minipage}{0.8\textwidth}
    \begin{mdframed}[
        linewidth=1pt,
        roundcorner=5pt
    ]
    \textsc{Monotone 1-in-3 SAT}
    
    \textbf{Input:} A 3-CNF formula $\varphi$ with no negation signs.

    \textbf{Output:}``Yes'': if $\varphi$ has a 1-in-3 assignment, under which each clause contains exactly one true literal;\\
    \hspace*{3.7em}``No'': otherwise.
    \end{mdframed}
    \end{minipage}
    \end{center}
    
    For example, $\varphi_1=(x_1 \vee x_2 \vee x_3) \wedge (x_2 \vee x_3 \vee x_4)$ is a yes-instance because $(x_1,x_2,x_3,x_4)=(\text{T,F,F,T})$ is an 1-in-3 assignment; $\varphi_2 = (x_1 \vee x_2 \vee x_3) \wedge (x_2 \vee x_3 \vee x_4) \wedge (x_1 \vee x_2 \vee x_4) \wedge (x_1 \vee x_3 \vee x_4)$ is a no-instance. 

    Given a 3-CNF formula $\varphi$, our goal is to construct a configuration of the training process, such that the resulting model outputs the correct answer if and only if its training data $T^\prime$ encodes an 1-in-3 assignment $\nu$ of $\varphi$. This can be done by carefully designing the encoding so that for each $x_i\in\varphi$, $\nu(x_i)=\textsc{True}$ if and only if $\mathbf{t}_{x_i}\in T^\prime$. Finally, we can construct some $T$ with $T\supseteq T^\prime\cup \{\mathbf{t}_{x_i}|x_i\in \varphi\}$, such that some classifier trained on $T$ is a yes-instance of \textsc{Debuggable-Lin} if and only if $\varphi$ is a yes-instance of \textsc{Monotone 1-in-3 SAT}, thereby finishing our proof.

    \textbf{The reduction.} Suppose $\varphi$ has $m$ clauses and $n$ variables, let $N=n+2m+1$. We set the dimension of the linear classifier to $N$.

    \underline{The input.} Each coordinate of the input is named as
    \begin{equation*}
        \mathbf{x}=(x_{c_1},\dots,x_{c_m},x_{x_1},\dots,x_{x_n},x_{b_1},\dots,x_{b_m},x_{\text{dummy}})^\top
    \end{equation*}
    We also use $x_i$ to denote the $i$-th coordinate of $\mathbf{x}$.

    \underline{The parameters.} Each coordinate of the parameter is named as
    \begin{equation*}
        \mathbf{w}=(w_{c_1},\dots,w_{c_m},w_{x_1},\dots,w_{x_n},w_{b_1},\dots,w_{b_m},w_{\text{dummy}})^\top
    \end{equation*} 
    We also use $w_i$ to denote the $i$-th coordinate of $\mathbf{w}$.
    Each $w_{x_j}$ represents the truth value of variable $x_j$, where 1 represents \textsc{True} and -1 represents \textsc{False}. 
    Similarly, each $w_{c_j}$ represents the truth value of clause $c_j$ based on the value of its variables. 
    $w_{b_j}$ and $w_\texttt{dummy}$ are used for convenience of proof.
    
    The initial value of the parameter is set to 
    \begin{equation*}
        \mathbf{w}^{(0)} = (\overbrace{\frac{1}{2},\dots,\frac{1}{2}}^{m},\overbrace{-1,\dots,-1}^{n},\overbrace{-1,\dots,-1}^{m},1)^\top
    \end{equation*}
    
    \underline{Loss function.} We denote $U(x_0,\delta):=\{x|x_0-\delta < x < x_0+\delta\}$ as the $\delta$-neighborhood of $x_0$ and define $U(\pm x_0,\delta)=U(x_0,\delta)\cup U(-x_0,\delta)$. We define the \textit{local ramp function} as 
    \begin{equation*}
        r_{x_0,\delta}(x)=
        \begin{cases}
            0 &, x\le x_0 - \delta;\\
            x-x_0 + \delta &, x\in U(x_0,\delta);\\
            2\delta &, x\ge x_0 + \delta.
        \end{cases}
    \end{equation*}
    
    The loss function is defined as
    \begin{equation*}
        \mathcal{L}= -\frac{12N}{5} r_{-5,0.01}(y\mathbf{w}^\top\mathbf{x}) - 
        r_{-\frac{1}{2},0.26}(y\mathbf{w}^\top\mathbf{x}) -
        \frac{1}{1000N}\sum_{x_0\in\{\pm 1,\pm 3\}}r_{x_0,0.01}(y\mathbf{w}^\top\mathbf{x}).
    \end{equation*}
    $\mathcal{L}$ is monotonically decreasing with derivatives
    \begin{equation}\label{gd-loss-derivative}
        \frac{\partial \mathcal{L}}{\partial w_i} =
        \begin{cases}
        -\frac{12N}{5}\cdot yx_i ~&, y\mathbf{w}^\top\mathbf{x}\in U(-5,0.01); \\
        -yx_i ~&, y\mathbf{w}^\top\mathbf{x}\in U(-\frac{1}{2},0.26);\\
        -\frac{1}{1000N}yx_i ~&, y\mathbf{w}^\top\mathbf{x} \in 
        \bigcup_{x_0\in\{\pm 1,\pm 3\}} U(x_0,0.01);\\
        0 ~&, \text{otherwise.}
        \end{cases}
    \end{equation}

    \underline{Learning rate.} The learning rate for SGD is set to be
    \begin{equation*}
        \boldsymbol{\eta}=(\overbrace{5,\dots,5}^{m},\overbrace{\frac{1}{6N},\dots,\frac{1}{6N}}^{n},\overbrace{2000N,\dots,2000N}^{m},1)^\top.
    \end{equation*}

    \underline{Training data.} We define two gadgets, \texttt{var($i$)} and \texttt{clause($i,i_1,i_2,i_3$)}, as illustrated in Table \ref{gadget_var} and \ref{gadget_clause}. All the unspecified coordinates are set to zero. We use $T_0$ to denote the training data.
    \texttt{var($i$)} is contained in $T_0$ if and only if $x_i \in \varphi$, and \texttt{clause($i,i_1,i_2,i_3$)} is contained in $T_0$ if and only if $c_i = (x_{i_1}\vee x_{i_2}\vee x_{i_3})\in\varphi$.
    
    \begin{table}
    \centering
    \begin{minipage}[t]{0.4\textwidth}
    \centering
    \caption{Training data for \texttt{var($i$)}}\label{gadget_var}
        \begin{tabular}{cc}
        \toprule
        $x_{x_i}$     &  y \\
        \midrule
        $5$ & $1$ \\
        \bottomrule
        \end{tabular}
    \end{minipage}
    \begin{minipage}[t]{0.5\textwidth}
        \centering
        \caption{Training data for \texttt{clause($i,i_1,i_2,i_3$)}}\label{gadget_clause}
        \begin{tabular}{cccccc}
        \toprule
        $x_{c_i}$ & $x_{x_{i_1}}$ & $x_{x_{i_2}}$ & $x_{x_{i_3}}$ & $x_{b_i}$ & y \\
        \midrule
        $1$ & $1$& $1$& $1$& $\frac{1}{2}$& $1$     \\
        \bottomrule
        \end{tabular}
    \end{minipage}   
    \end{table}

    \underline{Threshold and instance.} The threshold $\varepsilon$ can be any fixed value in $\mathbb{R}_{+}$. The instance is defined as $(\mathbf{x}_\text{test},y_\text{test})$, where $y_\text{test}=1$ and
    \begin{equation*}
        \mathbf{x}_\text{test}=(\overbrace{1,\dots,1}^{m},\overbrace{0,\dots,0}^{n+m},\frac{-11m+5}{2})^\top.
    \end{equation*}
    
    The following reduction works for all possible training orders. Intuitively, during the training process, each \texttt{var($i$)} in the training data will set $w_{x_i}$ to around $1$ (that is, mark $x_i$ as \textsc{True}) in the first epoch, and each \texttt{clause($i,i_1,i_2,i_3$)} will set $w_{c_i}$ to near $\frac{11}{2}$ in the second epoch, if and only if exactly one of $w_{x_{i_1}},w_{x_{i_2}},w_{x_{i_3}}$ is near $1$ and the others near $-1$ (that is,  mark $c_i$ as satisfied if exactly one of its literals is \textsc{True} and the others \textsc{False}). The training process terminates at the end of the second epoch.
    \end{proof}

\section{Results for Fixed Loss Functions}\label{fixed-loss}
  We have proved the NP-hardness for \textsc{Debuggable-Lin} when the loss function is not fixed. In this section, we study the complexity when the loss function is fixed as linear and hinge-like functions.   
  Assuming that SGD terminates after only one epoch with a fixed order, we will show that \textsc{Debuggable-Lin} is solvable in linear time for linear loss. For hinge-like loss functions, \textsc{Debuggable-Lin} can be solved in linear time only when the dimension $d=1$ and the interception $\beta\ge 0$. For the rest cases, \textsc{Debuggable-Lin} becomes NP-hard. 

  \subsection{The Easy Case}
    We start with the linear loss function $\mathcal{L}=-\alpha (y\mathbf{w}^\top\mathbf{x}+\beta)$, with which all the training data are activated and $\mathbf{w}^* = \mathbf{w}^*(T)=\mathbf{w}^{(0)} + \sum_{(\mathbf{x},y)\in T}\alpha y \boldsymbol{\eta}\otimes \mathbf{x}$.
    Since $y_\text{test}\in \{-1,1\}$, \textsc{Debuggable-Lin} is equivalent to deciding whether 
    \begin{equation*}
        \max_{T^\prime\subseteq T}\{y_\text{test} (\mathbf{w}^*(T^\prime))^\top \mathbf{x}_\text{test}\}>0.
    \end{equation*}
     
    A training sample $(\mathbf{x},y)$ is ``good'' if $y_\text{test} (\alpha y \boldsymbol{\eta}\otimes \mathbf{x})^\top \mathbf{x}_\text{test}>0$ and ``bad'' otherwise.
    The \emph{good training-sample assessment} (GTA) algorithm, as shown in Algorithm \ref{algo:gta}, deals with this situation by greedily picking all ``good'' training samples.
    

    Denoting $T^*$ as the set of all good data in $T$, it follows that
    \begin{equation*}
        \begin{aligned}
            y_\text{test}(\mathbf{w}^*(T^*))^\top \mathbf{x}_\text{test} &= y_\text{test}(\mathbf{w}^{(0)})^\top\mathbf{x}_\text{test} + \sum_{(\mathbf{x},y)\in T^*} y_\text{test} (\alpha y \boldsymbol{\eta}\otimes \mathbf{x})^\top \mathbf{x}_\text{test}\\
            &\ge y_\text{test}(\mathbf{w}^{(0)})^\top\mathbf{x}_\text{test} + \sum_{(\mathbf{x},y)\in T^\prime} y_\text{test} (\alpha y \boldsymbol{\eta}\otimes \mathbf{x})^\top \mathbf{x}_\text{test}
        \end{aligned}
    \end{equation*}
    for all $T^\prime\subseteq T$.
    Hence $\max_{T^\prime\subseteq T}\{y_\text{test} (\mathbf{w}^*(T^\prime))^\top \mathbf{x}_\text{test}\}=y_\text{test} (\mathbf{w}^*(T^*))^\top \mathbf{x}_\text{test}$ and \textsc{Debuggable-Lin} can be solved by GTA in linear time. 
    The following theorem is straightforward.
        
  \begin{theorem}\label{thm:linear-loss-linear-time}
      \textsc{Debuggable-Lin} is linear time solvable for linear loss functions.
  \end{theorem}

        

    \begin{algorithm}[H]
    \KwIn{Training data $T$, loss function 
    $\mathcal{L}$, initial parameter $\mathbf{w}^{(0)}$, learning rate $\boldsymbol{\eta}$, threshold $\varepsilon$ and test instance $(\mathbf{x}_\text{test},y_\text{test})$.
    }
    \KwOut{TRUE, iff $\texttt{SGD}_\Lambda(\mathcal{L},\boldsymbol{\eta},\varepsilon,T)$ is debuggable on $(\mathbf{x}_\text{test}y_\text{test})$.}
      \SetAlgoLined
      $\mathbf{w}\leftarrow \mathbf{w}^{(0)}$\; 
      \For{$(\mathbf{x},y)\in T$}{
        \If{$y_\text{test} (\alpha y \boldsymbol{\eta}\otimes \mathbf{x})^\top \mathbf{x}_\text{test}>0$}{
          $\mathbf{w}\leftarrow \mathbf{w}+\alpha y\boldsymbol{\eta}\otimes\mathbf{x}$\;}
      }
      \If{$y_\text{test} \mathbf{w}^\top \mathbf{x}_\text{test} \ge 0$}{return TRUE\;}
      return FALSE\;
      \caption{Good Training-sample Assessment (GTA)}\label{algo:gta}
    \end{algorithm}

    GTA is still effective for one-dimensional classifiers trained with hinge-like losses when $\beta\ge 0$. 
    \begin{theorem}\label{thm:hinge-loss-linear-time}
        \textsc{Debuggable-Lin} is linear time solvable for hinge-like loss functions, when $d=1$ and $\beta \ge 0$.
    \end{theorem}
    \begin{proof} 
        It suffices to prove that if $\exists T^\prime\subseteq T$ such that $\texttt{SGD}_\Lambda(T^\prime,x_\text{test})=y_\text{test} $, $\texttt{SGD}_\Lambda(T^*,x_\text{test})=y_\text{test}$.

        a) Suppose all the data in $T^*$ are activated, we have
        \begin{equation*}
            \begin{aligned}
                y_\text{test} w^*(T^*) x_\text{test} &= y_\text{test} w^{(0)} x_\text{test} + \sum_{(x,y)\in T^*}y_\text{test} \alpha y \eta x x_\text{test} \\
                &\ge y_\text{test} w^{(0)} x_\text{test} + \sum_{(x,y)\in T^\prime\cap T^*}y_\text{test} \alpha y \eta x x_\text{test} +\sum_{(x,y)\in T^\prime\setminus T^*}y_\text{test} \alpha y \eta x x_\text{test} \\
                &= y_\text{test} w^*(T^\prime) x_\text{test} \ge 0
            \end{aligned}
        \end{equation*}

        b) Suppose $(x,y)\in T^*$ is the first inactivated data during the training phase, and $w$ is the current parameter, we have $ywx>\beta$. Since $\alpha\eta \cdot (xy) \cdot (x_\text{test}y_\text{test}) \ge 0$, we have $(x_\text{test}y_\text{test})\cdot w\ge 0$.
        Let $T^{\prime\prime}$ be the set of training data appeared before $(x,y)$, we have $y_\text{test} w^*(T^*) x_\text{test} \ge y_\text{test} w^*(T^{\prime\prime}) x_\text{test}\ge 0$.
    \end{proof}

  \subsection{The Hard Case}

  The gradient of training data may not always be activated and could be affected by the training order. 
   When the training order is adversarially chosen, the following theorem shows that 
    \textsc{Debuggable-Lin} is NP-hard for all $d\ge 2$ and $\beta \in \mathbb{R}$.

  \begin{theorem}\label{thm:hinge-hard}
        If the training order is adversarially chosen and $d \ge 2$, \textsc{Debuggable-Lin} is NP-hard for \emph{each} hinge-like loss function at \emph{every} constant learning rate.
  \end{theorem}
  
  \begin{proof}[Proof sketch.]
   Since the result can be easily extended for all $d>2$ by padding the other $d-2$ dimensions with zeros, we only prove for the case of $d=2$.
    We assume $\beta \ge -1$ and leave the $\beta< -1$ case to the appendix. To avoid cluttering, we further assume $\boldsymbol{\eta}=\mathbf{1}$ and $\alpha=1$. The proof can be easily generalized by appropriately re-scaling the constructed vectors.
    
    We build a reduction from the subset sum problem, which is well-known to be NP-hard:
    \begin{center}
    \begin{minipage}{0.8\textwidth}
    \begin{mdframed}[
        linewidth=1pt,
        roundcorner=5pt
    ]
    \textsc{Subset Sum}
    
    \textbf{Input:} A set of positive integer $S$, and a positive integer $t$.

    \textbf{Output:} ``Yes'': if $\exists S^\prime\subseteq S$ such that $\sum_{a\in S^\prime}a = t$;\\
    \hspace*{3.7em}``No'': otherwise.
    \end{mdframed}
    \end{minipage}
    \end{center}

    Suppose $n=|S|$, $m=\max_{a\in S}\{a\}$, $\gamma = \max\{\beta,1\}$ and $S=\{a_1,a_2,\dots,a_n\}$. We further assume $n>1$.
    Let the training data be
    \begin{equation*}
        T=\{(\mathbf{x}_1,y_1),(\mathbf{x}_2,y_2),\dots,(\mathbf{x}_n,y_n)\}\cup\{(\mathbf{x}_c,y_c),(\mathbf{x}_b,y_b),(\mathbf{x}_a,y_a)\}
    \end{equation*}
    where $\mathbf{x}_iy_i=(\frac{\sqrt{\gamma}}{n+1},3\sqrt{\gamma} a_i) $ for all $1\le i\le n, \mathbf{x}_cy_c=((18n^2m^2-2)\sqrt{\gamma},-3t\sqrt{\gamma}),\mathbf{x}_by_b=(\sqrt{\gamma},-\sqrt{\gamma}), \mathbf{x}_ay_a=(\sqrt{\gamma},\sqrt{\gamma})$. Let $\mathbf{w}^{(0)}=(-18n^2m^2\sqrt{\gamma},0)$.
    Let the test instance $(\mathbf{x}_\text{test},y_\text{test})$ satisfy $\mathbf{x}_\text{test}y_\text{test}=(1,0)$. 
    
    Let the training order be $(\mathbf{x}_1,y_1),(\mathbf{x}_2,y_2),\dots,(\mathbf{x}_n,y_n),(\mathbf{x}_c,y_c),(\mathbf{x}_b,y_b),(\mathbf{x}_a,y_a)$.

    For each $1\le i<n$, suppose $\mathbf{w}^{(0)} \xrightarrow{T\cap \{(\mathbf{x}_i,y_i)|1\le j\le i\}}\mathbf{w}^{(i)}$, we have
      \begin{equation*}
          \begin{aligned}
              y_{i+1}(\mathbf{w}^{(i)})^\top\mathbf{x}_{i+1} &\le \frac{\sqrt{\gamma}}{n+1}(-18n^2m^2\sqrt{\gamma}+\frac{\sqrt{\gamma}i}{n+1}) + 3\sqrt{\gamma}a_{i+1}\sum_{j=1}^i 3\sqrt{\gamma}a_j\\
              &\le \gamma\left( -\frac{n-1}{n+1}\cdot 9nm^2+\frac{n}{(n+1)^2}\right) < -1 \le \beta
          \end{aligned}
      \end{equation*}
      
      This means all the $T\setminus \{(\mathbf{x}_c,y_c),(\mathbf{x}_b,y_b), (\mathbf{x}_a,y_a)\}$ can be activated. Thus the resulting parameter trained by $T\setminus \{(\mathbf{x}_c,y_c),(\mathbf{x}_b,y_b), (\mathbf{x}_a,y_a)\}$ is
      \begin{equation*}
          \mathbf{w}_c = \mathbf{w}^{(0)} + \sum_{i=1}^n \mathbf{x}_iy_i
          =\left( -18n^2m^2\sqrt{\gamma}+\frac{\sqrt{\gamma}|T^*|}{n+1} , 3\sqrt{\gamma} \sum_{i=1}^n a_i\right).
      \end{equation*}

    It now suffices to prove that for all $S^\prime\subseteq S$, $\sum_{a\in S^\prime} a = t$ if and only if $\exists T^\prime\subseteq T$ such that $\mathbf{w} : \mathbf{w}^{(0)} \xrightarrow{T^\prime}\mathbf{w}$ satisfies $y_\text{test}\mathbf{w}^\top\mathbf{x}_\text{test}>0$.

    \underline{If:} Suppose $\exists S^\prime\subseteq S$ such that $\sum_{a\in S}a=t$, we prove that $\exists T^\prime\subseteq T$ such that $y_\text{test}(\mathbf{w}^*)^\top\mathbf{x}_\text{test}>0$ for $\mathbf{w}^*$ satisfying $\mathbf{w}^{(0)}\xrightarrow{T^\prime}\mathbf{w}^*$.
    
    Let $T^*=\{(\mathbf{x}_i,y_i)|a_i\in S^\prime\}$, $T^\prime = T^*\cup 
    \{(\mathbf{x}_c,y_c),(\mathbf{x}_b,y_b), (\mathbf{x}_a,y_a)\}$. We have
    \begin{equation*}
        \mathbf{w}_c = (-18n^2m^2\sqrt{\gamma} + \frac{\sqrt{\gamma}|T^*|}{n+1},3\sqrt{\gamma}\sum_{a_i\in S^\prime}a_i)= (-18n^2m^2\sqrt{\gamma} + \frac{\sqrt{\gamma}|T^*|}{n+1},3\sqrt{\gamma}t).
    \end{equation*}
    And therefore $y_c\mathbf{w}_c^\top\mathbf{x}_c=\gamma\left((-18n^2m^2+\frac{|T^*|}{n+1})(18n^2m^2-2) -9t^2 \right)<-1\le \beta$, so
    \begin{equation*}
        \mathbf{w}_c \xrightarrow{(\mathbf{x}_c,y_c)}\mathbf{w}_b = \mathbf{w}_c + \mathbf{x}_cy_c = (\sqrt{\gamma}(\frac{|T^*|}{n+1}-2),0).
    \end{equation*}
    Note that $y_b\mathbf{w}_b^\top\mathbf{x}_b=\gamma(\frac{|T^*|}{n+1}-2)< -1 \le \beta$, we have
    \begin{equation*}
        \mathbf{w}_b \xrightarrow{(\mathbf{x}_b,y_b)}\mathbf{w}_a = \mathbf{w}_b + \mathbf{x}_ay_a = (\sqrt{\gamma}(\frac{|T^*|}{n+1}-1),-\sqrt{\gamma})
    \end{equation*}
    Note also that $y_a\mathbf{w}_a^\top\mathbf{x}_a=\gamma(\frac{|T^*|}{n+1}-2)<-1\le \beta$, we have
    \begin{equation*}
        \mathbf{w}_a \xrightarrow{(\mathbf{x}_a,y_a)}\mathbf{w}^* = \mathbf{w}_a + \mathbf{x}_ay_a = (\frac{|T^*|\sqrt{\gamma}}{n+1},0)
    \end{equation*}
    Therefore, $y_\text{test}(\mathbf{w}^*)^\top\mathbf{x}_\text{test}=\frac{|T^*|\sqrt{\gamma}}{n+1}>0$.

    \underline{Only if:}
    For each $T^\prime\subseteq T$, let $T^*=T'\setminus \{(\mathbf{x}_c,y_c),(\mathbf{x}_b,y_b), (\mathbf{x}_a,y_a)\}$.
    If $y_\text{test}(\mathbf{w}^*)^\top\mathbf{x}_\text{test}>0$ for $\mathbf{w}^*$ satisfying $\mathbf{w}^{(0)}\xrightarrow{T^\prime}\mathbf{w}^*$, we prove that $\exists S^\prime\subseteq S$ such that $\sum_{a\in S^\prime}a=t$. 
    We first show that for each $T^\prime\subseteq T$, if $\mathbf{w}(\mathbf{w}^{(0)}\xrightarrow{T^\prime}\mathbf{w})$ satisfying $y_\text{test}\mathbf{w}^\top\mathbf{x}_\text{test}>0$, we have $\forall k\in\{a,b,c\},(\mathbf{x}_k,y_k)\in T^\prime,\warning{y_k\mathbf{w}_k^\top\mathbf{x}_k}<\gamma$, where $\mathbf{w}^{(0)}\xrightarrow{T^*}\mathbf{w}_c\xrightarrow{(\mathbf{x}_c,y_c)}\mathbf{w}_b\xrightarrow{(\mathbf{x}_b,y_b)}\mathbf{w}_a$. Otherwise, suppose $\exists k \in \{a,b,c\}$ such that $(\mathbf{x}_k,y_k)\not\in T^\prime$ or $y_k\mathbf{w}_k^\top\mathbf{x}_k\ge\gamma$, we have
    \begin{equation*}
    \begin{aligned}
        y_\text{test}\mathbf{w}^\top\mathbf{x}_\text{test} \le 
        \sqrt{\gamma}(\frac{|T^*|}{n+1}-1)<0
    \end{aligned}
    \end{equation*}
    which contradicts to the fact that $y_\text{test}\mathbf{w}^\top\mathbf{x}_\text{test}\ge 0$.

    Let $S^\prime=\{a_i|(\mathbf{x}_i,y_i)\in T^*\}$ and $t^\prime=\sum_{a\in S^\prime}a_i$, it suffices to prove $t^\prime=t$.
    Notice that 
    \begin{equation*}
    \begin{aligned}
        \mathbf{w}^{(0)}\xrightarrow{T^*\cap \{(\mathbf{x}_i,y_i)|1\le j\le i\}}\mathbf{w}_c &= (\sqrt{\gamma}(-18n^2m^2+\frac{|T^*|}{n+1}),3\sqrt{\gamma}\sum_{a_i\in S^\prime}a_i)\\
        &=(\sqrt{\gamma}(-18n^2m^2+\frac{|T^*|}{n+1}),3\sqrt{\gamma}t^\prime)
    \end{aligned}
    \end{equation*}
    Hence $y_c\mathbf{w}_c^\top\mathbf{x}_c=\gamma(-18n^2m^2+\frac{|T^*|}{n+1})(18n^2m^2-2)-9\gamma tt^\prime<-1\le\beta$, thus
    \begin{equation*}
        \mathbf{w}_c\xrightarrow{(\mathbf{x}_c,y_c)}\mathbf{w}_b=\mathbf{w}_c + \mathbf{x}_cy_c = (\sqrt{\gamma}(\frac{|T^*|}{n+1}-2),3\sqrt{\gamma}(t^\prime-t))
    \end{equation*}

    (1) If $t^\prime\le t-1$, we have $y_b\mathbf{w}_b^\top\mathbf{x}_b=\gamma \left(\frac{|T^*|}{n+1}-2+3(t-t^\prime)\right)> \gamma\ge \beta$, a contradiction.

    (2) If $t^\prime\ge t+1$, we have $y_a\mathbf{w}_a^\top\mathbf{x}_a=\gamma \left(\frac{|T^*|}{n+1}-2+3(t^\prime-t)\right)> \gamma\ge\beta$, another contradiction.

    Therefore $t^\prime = t$, and this completes the proof.
  \end{proof}

    Moreover, \textsc{Debuggable-Lin} is NP-hard even when $d=1$ and $\beta<0$.
   \begin{theorem}\label{1d-hinge-hard}
       If the training order is adversarially chosen and $d = 1$, \textsc{Debuggable-Lin} remains NP-hard for \emph{each} hinge-like loss function with $\beta<0$ at \emph{every} constant learning rate.
   \end{theorem}
  
 \textbf{Remarks.} The training order in this section can be arbitrary as long as the last three training samples are $(\mathbf{x}_c,y_c),(\mathbf{x}_b,y_b),(\mathbf{x}_a,y_a)$, respectively. All the training samples are ``good'' since for each $(\mathbf{x},y)\in T$ we have $\mathbf{x}^\top\mathbf{x}_\text{test}yy_\text{test}>0$. This implies that \textsc{Debuggable-Lin} is NP-hard even if all the training data are ``good'' training samples, and exemplifies why the GTA algorithm fails for higher dimensions.
    
\section{Discussion and Conclusion}\label{sec-discussion}

In this paper, we provided a comprehensive analysis on the complexity of \textsc{Debuggable}.
We focus on the linear classifier that is trained using SGD, as it is a key component in the majority of popular models.

Since \textsc{Debuggable} is a special case of data debugging, the above results proved the intractability of data debugging and therefore gives a negative answer to Problem \ref{open-problem} declared in the introduction. The complexity results also demonstrated that it is not accurate to estimate the impact of subset of training data by summing up the score of each training samples in the subset, \emph{as long as the scores can be calculated in polynomial time}.

In Section \ref{fixed-loss}, a training sample is said to be ``good'' if it can help the resulting model to predict correctly on the test instance. That is, it can increase $y_\text{test}(\mathbf{w}^*)^\top\mathbf{x}_\text{test}$. 
However, in our proof we showed that \textsc{Debuggable} remains NP-hard even if all training samples are ``good''. This suggests that the quality of a training sample does not depend only on some properties of itself but also on the interaction between the rest of the training data, which should be taken into consideration when developing data cleaning approaches.

Moreover, the NP-hardness of \textsc{Debuggable}  implies that, it is in general intractable to figure out the causality between even the prediction of a linear classifier and its training data. This may be seem surprising since linear classifiers have long been considered ``inherently interpretable''. As warned in  \cite{Jacovi2020TowardsFI}, \textit{a method being ``inherently interpretable'' needs to be verified before it can be trusted}, the concept of interpretability must be \textit{rigorously defined}, or at least its boundaries specified.

Our results suggests the following directions for future research. Firstly, characterizing the training sample may be helpful in designing efficient algorithms for data debugging; Secondly, designing algorithms using CSP-solver is a potential way to solve data debugging more efficiently than the brute-force algorithms; Finally, developing random algorithms is a potential way to solve data debugging successfully with high probability.

\bibliographystyle{plain}
\bibliography{reference}

\begin{thebibliography}{10}

\bibitem{influence-func-answer}
Juhan Bae, Nathan Ng, Alston Lo, Marzyeh Ghassemi, and Roger Grosse.
\newblock If influence functions are the answer, then what is the question?
\newblock In {\em Proceedings of the 36th International Conference on Neural Information Processing Systems}, NIPS '22, Red Hook, NY, USA, 2024. Curran Associates Inc.

\bibitem{SO-IF}
Samyadeep Basu, Xuchen You, and Soheil Feizi.
\newblock On second-order group influence functions for black-box predictions.
\newblock In {\em Proceedings of the 37th International Conference on Machine Learning}, ICML'20. JMLR.org, 2020.

\bibitem{brunet2019understanding}
Marc-Etienne Brunet, Colleen Alkalay-Houlihan, Ashton Anderson, and Richard Zemel.
\newblock Understanding the origins of bias in word embeddings, 2019.

\bibitem{order-of-training-sample}
Ernie Chang, Hui-Syuan Yeh, and Vera Demberg.
\newblock Does the order of training samples matter? improving neural data-to-text generation with curriculum learning.
\newblock {\em ArXiv}, abs/2102.03554, 2021.

\bibitem{Computability-book}
Erik~D. Demaine, William Gasarch, and Mohammad Hajiaghayi.
\newblock {\em Computational Intractability: A Guide to Algorithmic Lower Bounds}.
\newblock MIT Press, 2024.

\bibitem{Deng1994OnTC}
Xiaotie Deng and Christos~H. Papadimitriou.
\newblock On the complexity of cooperative solution concepts.
\newblock {\em Math. Oper. Res.}, 19:257--266, 1994.

\bibitem{Ghorbani2019DataSE}
Amirata Ghorbani and James~Y. Zou.
\newblock Data shapley: Equitable valuation of data for machine learning.
\newblock {\em ArXiv}, abs/1904.02868, 2019.

\bibitem{guo-etal-2021-fastif}
Han Guo, Nazneen Rajani, Peter Hase, Mohit Bansal, and Caiming Xiong.
\newblock {F}ast{IF}: Scalable influence functions for efficient model interpretation and debugging.
\newblock In Marie-Francine Moens, Xuanjing Huang, Lucia Specia, and Scott Wen-tau Yih, editors, {\em Proceedings of the 2021 Conference on Empirical Methods in Natural Language Processing}, pages 10333--10350, Online and Punta Cana, Dominican Republic, November 2021. Association for Computational Linguistics.

\bibitem{nips19-data-cleansing-with-sgd}
Satoshi Hara, Atsushi Nitanda, and Takanori Maehara.
\newblock {\em Data Cleansing for Models Trained with SGD}.
\newblock Curran Associates Inc., Red Hook, NY, USA, 2019.

\bibitem{Jacovi2020TowardsFI}
Alon Jacovi and Yoav Goldberg.
\newblock Towards faithfully interpretable nlp systems: How should we define and evaluate faithfulness?
\newblock In {\em Annual Meeting of the Association for Computational Linguistics}, 2020.

\bibitem{Jia2019TowardsED}
R.~Jia, David Dao, Boxin Wang, Frances~Ann Hubis, Nicholas Hynes, Nezihe~Merve G{\"u}rel, Bo~Li, Ce~Zhang, Dawn~Xiaodong Song, and Costas~J. Spanos.
\newblock Towards efficient data valuation based on the shapley value.
\newblock {\em ArXiv}, abs/1902.10275, 2019.

\bibitem{shap-knn-proxy}
Ruoxi Jia, David Dao, Boxin Wang, Frances~Ann Hubis, Nezihe~Merve Gurel, Bo~Li, Ce~Zhang, Costas Spanos, and Dawn Song.
\newblock Efficient task-specific data valuation for nearest neighbor algorithms.
\newblock {\em Proc. VLDB Endow.}, 12(11):1610–1623, jul 2019.

\bibitem{sca-util-shap}
Ruoxi Jia, Fan Wu, Xuehui Sun, Jiacen Xu, David Dao, Bhavya Kailkhura, Ce~Zhang, Bo~Li, and Dawn Song.
\newblock Scalability vs. utility: Do we have to sacrifice one for the other in data importance quantification?
\newblock In {\em 2021 IEEE/CVF Conference on Computer Vision and Pattern Recognition (CVPR)}, pages 8235--8243, 2021.

\bibitem{karlaš2022data-shapley}
Bojan Karlaš, David Dao, Matteo Interlandi, Bo~Li, Sebastian Schelter, Wentao Wu, and Ce~Zhang.
\newblock Data debugging with shapley importance over end-to-end machine learning pipelines, 2022.

\bibitem{Khanna2018InterpretingBB}
Rajiv Khanna, Been Kim, Joydeep Ghosh, and Oluwasanmi Koyejo.
\newblock Interpreting black box predictions using fisher kernels.
\newblock In {\em International Conference on Artificial Intelligence and Statistics}, 2018.

\bibitem{nips19-influence-liang}
Pang~Wei Koh, Kai-Siang Ang, Hubert Hua~Kian Teo, and Percy Liang.
\newblock On the accuracy of influence functions for measuring group effects.
\newblock In {\em Neural Information Processing Systems}, 2019.

\bibitem{ICML-influence}
Pang~Wei Koh and Percy Liang.
\newblock Understanding black-box predictions via influence functions.
\newblock In {\em Proceedings of the 34th International Conference on Machine Learning - Volume 70}, ICML'17, page 1885–1894. JMLR.org, 2017.

\bibitem{CleanML}
Peng Li, Xi~Rao, Jennifer Blase, Yue Zhang, Xu~Chu, and Ce~Zhang.
\newblock Cleanml: A study for evaluating the impact of data cleaning on ml classification tasks.
\newblock In {\em 2021 IEEE 37th International Conference on Data Engineering (ICDE)}, pages 13--24, 2021.

\bibitem{FROG}
Yejia Liu, Weiyuan Wu, Lampros Flokas, Jiannan Wang, and Eugene Wu.
\newblock Enabling sql-based training data debugging for federated learning.
\newblock {\em Proceedings of the VLDB Endowment}, 15:388--400, 02 2022.

\bibitem{training-data-order}
Jeremy Mange.
\newblock Effect of training data order for machine learning.
\newblock In {\em 2019 International Conference on Computational Science and Computational Intelligence (CSCI)}, pages 406--407, 2019.

\bibitem{Neutatz2021FromCB}
Felix Neutatz, Binger Chen, Ziawasch Abedjan, and Eugene Wu.
\newblock From cleaning before ml to cleaning for ml.
\newblock {\em IEEE Data Eng. Bull.}, 44:24--41, 2021.

\bibitem{subset}
Victor Parque.
\newblock Tackling the subset sum problem with fixed size using an integer representation scheme.
\newblock In {\em 2021 IEEE Congress on Evolutionary Computation (CEC)}, pages 1447--1453, 2021.

\bibitem{SIGMOD22-interp-data-based-exp}
Romila Pradhan, Jiongli Zhu, Boris Glavic, and Babak Salimi.
\newblock Interpretable data-based explanations for fairness debugging.
\newblock In {\em Proceedings of the 2022 International Conference on Management of Data}, SIGMOD '22, page 247–261, New York, NY, USA, 2022. Association for Computing Machinery.

\bibitem{wang2019repairing}
Hao Wang, Berk Ustun, and Flavio~P. Calmon.
\newblock Repairing without retraining: Avoiding disparate impact with counterfactual distributions, 2019.

\bibitem{loss-survey}
Qi~Wang, Yue Ma, Kun Zhao, and Yingjie Tian.
\newblock A comprehensive survey of loss functions in machine learning.
\newblock {\em Annals of Data Science}, 9, 04 2022.

\bibitem{Query2.0}
Weiyuan Wu, Lampros Flokas, Eugene Wu, and Jiannan Wang.
\newblock Complaint-driven training data debugging for query 2.0.
\newblock pages 1317--1334, 06 2020.

\end{thebibliography}

\newpage
\appendix
\section{Detailed Proofs for Section \ref{gd-hardness}}\label{gd-proof}
    \textbf{Notations.} Given some orderings $\{o_e\}$ of training data, where $o_{\mathbf{t}}^e$ as the order of $\mathbf{t}$ in epoch $e$. We use $w_{x_i}^{(e,l)}$ to denote the value of $w_{x_i}$ after the $l$-th iteration in epoch $e$. We also denote $\mathbf{x_t}$ and $y_\mathbf{t}$ as the feature and the label of training data $\mathbf{t}$, respectively. We denote $\mathbf{t}^{(e,l)}$ as the training sample being considered during epoch $e$, iteration $l$.

    \begin{lemma}\label{gd-weight-clause-missing}
        Suppose $T\subseteq T_0$ is the training data and let  $T^e_{l,r} =\{\mathbf{t}^{(e,l)}, \mathbf{t}^{(e,l+1)},\dots, \mathbf{t}^{(e,r)}\}$ be the set of consecutive training samples considered during epoch $e$ from iteration $l$ to $r$. For $1\le l\le r \le |T|$, if \texttt{clause}($\gamma,i_1,i_2,i_3$)$\not\in T^e_{l,r}$, then $w_{c_\gamma}^{(e,l-1)} = w_{c_\gamma}^{(e,r)}$.
    \end{lemma}
    \begin{proof}
        For each $\mathbf{t}\in T_{l,r}^e$, we have $ (x_\mathbf{t})_{c_\gamma}=0$. Therefore 
        \begin{equation*}
            \left| \left. \frac{\partial\mathcal{L}}{\partial c_\gamma} \right|_{\mathbf{t}} \right| \le \max\left\{\left|-\frac{12N}{5}yx_{c_\gamma}\right|,|-yx_{c_\gamma}|,\left|-\frac{1}{1000N}yx_{c_\gamma}\right|,0\right\} = 0
        \end{equation*}
        Hence $\left. \frac{\partial\mathcal{L}}{\partial c_\gamma} \right|_{\mathbf{t}}=0$, and
        \begin{equation*}
            w_{c_\gamma}^{(e,r)} = w_{c_\gamma}^{(e,l-1)} - \eta_{c_\gamma} \sum_{\mathbf{t}\in T_{l,r}^e}\left. \frac{\partial\mathcal{L}}{\partial c_\gamma} \right|_{\mathbf{t}}=w_{c_\gamma}^{(e,l-1)}
        \end{equation*}
        
        Similarly, $(x_\mathbf{t})_{b_\gamma}=0$, and
        \begin{equation*}
            \left| \left. \frac{\partial\mathcal{L}}{\partial b_\gamma} \right|_{\mathbf{t}} \right| \le \max\left\{\left|-\frac{12N}{5}yx_{b_\gamma}\right|,|-yx_{b_\gamma}|,\left|-\frac{1}{1000N}yx_{b_\gamma}\right|,0\right\} = 0
        \end{equation*}
        Hence $\left. \frac{\partial\mathcal{L}}{\partial b_\gamma} \right|_{\mathbf{t}}=0$, and
        
        \begin{equation*}
            w_{b_\gamma}^{(e,r)} = w_{b_\gamma}^{(e,l-1)} - \eta_{b_\gamma} \sum_{\mathbf{t}\in T_{l,r}^e}\left. \frac{\partial\mathcal{L}}{\partial b_\gamma} \right|_{\mathbf{t}}=w_{b_\gamma}^{(e,l-1)}
        \end{equation*}
    \end{proof}

    \begin{lemma}\label{gd-weight-epoch1-val-iter}
        Suppose $T\subseteq T_0$ is the training data and $T_l := \{\mathbf{t}^{(1,1)},\dots,\mathbf{t}^{(1,l)}\}$. 
        $\forall 1\le i\le n, 1\le l \le |T|$, $w_{x_i}^{(1,l)}\in U(1,\frac{l+1}{6000N^2})$ if \texttt{var($i$)}$\in T_l$; Otherwise $w_{x_i}^{(1,l)}\in U(-1,\frac{l+1}{6000N^2})$.
    \end{lemma}
    \begin{proof}
        We prove this lemma by induction.
        
        \underline{Basic Case:} Note that for all $1\le i\le n$, $w_{x_i}^{(0)}=-1$, and for all $1\le \gamma \le m, w_{c_\gamma}^{(0)}=1/2, w_{b_\gamma}^{(0)}=-1$.
        We denote $\mathbf{t}=\mathbf{t}^{(1,1)}$ to avoid cluttering. For any fixed $i$:

        (1) If $\mathbf{t}=$\texttt{var($i$)}. We have $y_\mathbf{t} (\mathbf{w}^{(0)})^\top \mathbf{x}_\mathbf{t}^\prime = 5w_{x_i}^{(0)}=-5$, hence 
        \begin{equation*}
            \left.\frac{\partial\mathcal{L}}{\partial w_{x_i}}\right|_\mathbf{t} = -\frac{12N}{5}y_\mathbf{t}(x_\mathbf{t})_i=-12N
        \end{equation*} and 
        \begin{equation*}
            w_{x_i}^{(1,1)}=w_{x_i}^{(0)} - \eta_{x_i}\left.\frac{\partial\mathcal{L}}{\partial w_{x_i}}\right|_\mathbf{t} = -1 - \frac{1}{6N}\left( -\frac{12N}{5}\right) = 1 \in U(1,\frac{2}{6000N^2})
        \end{equation*}

        (2) If $\mathbf{t}=$\texttt{clause($\gamma,i,i^\prime,i^{\prime\prime}$)}. 
        We have
        \begin{equation*}
            y_\mathbf{t} (\mathbf{w}^{(0)})^\top \mathbf{x}_\mathbf{t}^\prime = w_{x_i}^{(0)}+ w_{x_{i^\prime}}^{(0)}+ w_{x_{i^{\prime\prime}}}^{(0)} + w_{c_\gamma}^{(0)} + \frac{1}{2}w_{b_\gamma}^{(0)}= -3
        \end{equation*}
        hence
        \begin{equation*}
            \left.\frac{\partial\mathcal{L}}{\partial w_{x_i}}\right|_\mathbf{t} = -\frac{1}{1000N}y_\mathbf{t}(x_\mathbf{t})_{x_i}=-\frac{1}{1000N}
        \end{equation*} and 
        \begin{equation*}
        \begin{aligned}
             w_{x_i}^{(1,1)}&=w_{x_i}^{(0)} - \eta_{x_i}\left.\frac{\partial\mathcal{L}}{\partial w_{x_i}}\right|_\mathbf{t}= -1 - \frac{1}{6N}\left( -\frac{1}{1000N}\right) \\
             &= -1 + \frac{1}{6000N^2} \in U(-1,\frac{2}{6000N^2})
        \end{aligned}
        \end{equation*}
        
        (3) Otherwise, $w_{x_i}$ will not be updated. Therefore $w_{x_i}^{(1,1)}=w_{x_i}^{(0)}=-1 \in U(-1,\frac{2}{6000N^2})$.

        Hence this lemma is true for $l=1$.

        \underline{Induction Step:} Suppose the lemma is true for $l<|T|$. We prove that this lemma remains true for $l+1$. We denote $\mathbf{t}=\mathbf{t}^{(1,l+1)}$ to avoid cluttering. This makes sense since $l+1\le |T|$ and thus $\mathbf{t}\in T$. For any fixed $i$:
        
        (1) If $\mathbf{t}=$\texttt{var($i$)}, then \texttt{var($i$)}$\not\in T_{l}$ because there are at most one \texttt{var($i$)} in $T$ for each $i$. 
        
        Therefore $w_{x_i}^{(1,l)}\in U(-1,\frac{l+1}{6000N^2})$.
        We have $y_\mathbf{t} (\mathbf{w}^{(1,l)})^\top \mathbf{x}_\mathbf{t}^\prime = 5w_{x_i}^{(1,l)}\in U(-5,0.01)$, and $\left.\frac{\partial\mathcal{L}}{\partial w_{x_i}}\right|_\mathbf{t} = -\frac{12N}{5}y_\mathbf{t}(x_\mathbf{t})_i=-12N$.
        Hence 
        \begin{equation*}
            \begin{aligned}
                w_{x_i}^{(1,l+1)}&=w_{x_i}^{(1,l)}  -\eta_{x_i}\left.\frac{\partial\mathcal{L}}{\partial w_{x_i}}\right|_\mathbf{t} = w_{x_i}^{(1,l)} - \frac{1}{6N}\left( -\frac{12N}{5}\right) \\
                &= w_{x_i}^{(1,l)} +2 \in U(1,\frac{l+2}{6000N^2})
            \end{aligned}
        \end{equation*}
        
        (2) If $\mathbf{t}=$\texttt{clause($\gamma,i,i^\prime,i^{\prime\prime}$)}. 
        In this case, \texttt{clause($\gamma,\cdot,\cdot,\cdot$)}$\not\in T_{1,l}^1$ and by Lemma \ref{gd-weight-clause-missing} we have $w_{c_\gamma}^{(1,l)}=w_{c_\gamma}^{(0)}, w_{b_\gamma}^{(1,l)}=w_{b_\gamma}^{(0)}$.    
        From the induction hypothesis we have 
        \begin{equation*}
            w_{x_i}^{(1,l)},w_{x_{i^\prime}}^{(1,l)},w_{x_{i^{\prime\prime}}}^{(1,l)}\in U(\pm 1, \frac{l+1}{6000N^2})
        \end{equation*}
        and thus 
        \begin{equation*}
            \begin{aligned}
                y_\mathbf{t} (\mathbf{w}^{(1,l)})^\top \mathbf{x}_\mathbf{t}^\prime &= w_{x_i}^{(1,l)}+ w_{x_{i^\prime}}^{(1,l)}+ w_{x_{i^{\prime\prime}}}^{(1,l)} + w_{c_\gamma}^{(1,l)} + \frac{1}{2}w_{b_\gamma}^{(1,l)}\\
                &= w_{x_i}^{(1,l)}+ w_{x_{i^\prime}}^{(1,l)}+ w_{x_{i^{\prime\prime}}}^{(1,l)}\\
                &\in \bigcup_{x_0 \in \{\pm 1,\pm 3 \}} U(x_0, \frac{3(l+1)}{6000N^2}) \subseteq \bigcup_{x_0 \in \{\pm 1,\pm 3 \}} U(x_0, 0.01)
            \end{aligned}
        \end{equation*}
        We have $\left.\frac{\partial\mathcal{L}}{\partial w_{x_i}}\right|_\mathbf{t} = -\frac{1}{1000N}$ and $w_{x_i}^{(1,l+1)} = w_{x_i}^{(1,l)}-\eta_{x_i}\left.\frac{\partial\mathcal{L}}{\partial w_{x_i}}\right|_\mathbf{t}=w_{x_i}^{(1,l)} + \frac{1}{6000N^2}$. Consider the following cases:

        \begin{itemize}
            \item If \texttt{var}($i$)$\in T_l$, then \texttt{var}($i$)$\in T_{l+1}$ and $w_{x_i}^{(1,l)}\in U(1,\frac{l+1}{6000N^2})$. Therefore $w_{x_i}^{(1,l+1)}\in U(1,\frac{l+2}{6000N^2})$.
            \item If \texttt{var}($i$)$\not\in T_l$, then \texttt{var}($i$)$\not\in T_{l+1}$ and $w_{x_i}^{(1,l)}\in U(-1,\frac{l+1}{6000N^2})$. Therefore $w_{x_i}^{(1,l+1)}\in U(-1,\frac{l+2}{6000N^2})$.
        \end{itemize}

        (3) Otherwise, $w_{x_i}$ will not be updated, and $w_{x_i}^{(1,l+1)}=w_{x_i}^{(1,l)}$. If \texttt{var}($i$)$\in T_l$ then \texttt{var}($i$)$\in T_{l+1}$ and $w_{x_i}^{(1,l+1)}\in U(1,\frac{l+2}{6000N^2})$; Otherwise \texttt{var}($i$)$\not\in T_{l+1}$ and $w_{x_i}^{(1,l+1)}\in U(-1,\frac{l+2}{6000N^2})$.

        Hence if the lemma is true for $l<|T|$, it is also true for $l+1$.
        Therefore, the lemma is true for all $1\le l\le |T|$.
    \end{proof}

    \begin{corollary}\label{gd-weight-epoch1-val}
        Suppose $T\subseteq T_0$ is the training data. 
        $\forall 1\le i\le n, 1\le l \le |T|$, if \texttt{var($i$)}$\in T$, then $w_{x_i}^{(1)}\in U(1, \frac{1}{6000N})$. Otherwise $w_{x_i}^{(1)}\in U(-1, \frac{1}{6000N})$.
    \end{corollary}
    \begin{proof}
        Note that $w_{x_i}^{(1)} = w_{x_i}^{(1,|T|)}$ and $N=2m+n+1$. By Lemma \ref{gd-weight-epoch1-val-iter}, if \texttt{var($i$)}$\in T$ we have 
        \begin{equation*}
            w_{x_i}^{(1,|T|)}\in U(1,\frac{|T|+1}{6000N^2}) \subseteq U(1,\frac{m+n+1}{6000N^2})\subseteq U(1,\frac{1}{6000N})
        \end{equation*}
         If \texttt{var($i$)}$\not\in T$, we have 
         \begin{equation*}
            w_{x_i}^{(1,|T|)}\in U(-1,\frac{|T|+1}{6000N^2})\subseteq U(-1,\frac{m+n+1}{6000N^2}) \subseteq U(-1,\frac{1}{6000N})    
         \end{equation*}
    \end{proof}
    
    \begin{lemma}\label{gd-weight-epoch1-clause} Suppose $T\subseteq T_0$ is the training data. 
    $\forall 1\le \gamma\le m$, if $\exists 1\le i_1,i_2,i_3 \le n$ such that \texttt{clause($\gamma,i_1,i_2,i_3$)} $\in T$, then $w_{b_\gamma}^{(1)}=0, w_{c_\gamma}^{(1)} = \frac{1}{2} + \frac{1}{200N}$;
    Otherwise, $w_{b_\gamma}^{(1)}=-1, w_{c_\gamma}^{(1)} =\frac{1}{2}$.
    \end{lemma}

    \begin{proof}
        (1) If such $\mathbf{t}_\gamma=$\texttt{clause($\gamma,i_1,i_2,i_3$)} exists in $T$, by Lemma \ref{gd-weight-epoch1-val-iter} we have 
        \begin{equation*}
        w_{x_{i_1}}^{(1,o_{\mathbf{t}_\gamma}^1)}+w_{x_{i_2}}^{(1,o_{\mathbf{t}_\gamma}^1)}+w_{x_{i_3}}^{(1,o_{\mathbf{t}_\gamma}^1)}\in \bigcup_{x_0 \in \{\pm 1,\pm 3\}}U(x_0,\frac{3(o_{\mathbf{t}_\gamma}^1+1)}{6000N^2})\subseteq \bigcup_{x_0 \in \{\pm 1,\pm 3\}}U(x_0,0.01)
        \end{equation*}
        
        By Lemma \ref{gd-weight-clause-missing} we have $w_{c_\gamma}^{(1,o_{\mathbf{t}_\gamma}^1-1)}=w_{c_\gamma}^{(0)}$ and $w_{b_\gamma}^{(1,o_{\mathbf{t}_\gamma}^1-1)}=w_{b_\gamma}^{(0)}$ because \texttt{clause($\gamma,\cdot,\cdot,\cdot$)}$\not\in T^1_{1,o_{\mathbf{t}_\gamma}-1}$.
        Hence
        \begin{equation*}
        \begin{aligned}
             y_{\mathbf{t}_\gamma}(\mathbf{w}^{(1,o_{\mathbf{t}_\gamma}^1-1)})^\top\mathbf{x}_{\mathbf{t}_\gamma}^\prime &= 
            w_{x_{i_1}}^{(1,o_{\mathbf{t}_\gamma}^1)}+w_{x_{i_2}}^{(1,o_{\mathbf{t}_\gamma}^1)}+w_{x_{i_3}}^{(1,o_{\mathbf{t}_\gamma}^1)} + w_{c_\gamma}^{(1,o_{\mathbf{t}_\gamma}^1-1)} + \frac{1}{2}w_{b_\gamma}^{(1,o_{\mathbf{t}_\gamma}^1-1)} \\
            &=w_{x_{i_1}}^{(1,o_{\mathbf{t}_\gamma}^1)}+w_{x_{i_2}}^{(1,o_{\mathbf{t}_\gamma}^1)}+w_{x_{i_3}}^{(1,o_{\mathbf{t}_\gamma}^1)} + w_{c_\gamma}^{(1,o_{\mathbf{t}_\gamma}^1-1)} \\            
            &\in \bigcup_{x_0 \in \{\pm 1,\pm 3\}}U(x_0,0.01)
        \end{aligned}
        \end{equation*}
        We have $\left. \frac{\partial\mathcal{L}}{\partial w_{c_\gamma}} \right|_{\mathbf{t}_\gamma} = -\frac{1}{1000N}$, and
        \begin{equation*}
            w_{c_\gamma}^{(1,o_{\mathbf{t}_\gamma}^1)} = w_{c_\gamma}^{(1,o_{\mathbf{t}_\gamma}^1-1)} - \eta_{c_\gamma}
            \left. \frac{\partial\mathcal{L}}{\partial w_{c_\gamma}} \right|_{\mathbf{t}_\gamma} = \frac{1}{2} +5 \times \frac{1}{1000N}=\frac{1}{2}+\frac{1}{200N}
        \end{equation*}
        Similarly, $\left. \frac{\partial\mathcal{L}}{\partial w_{b_\gamma}} \right|_{\mathbf{t}_\gamma} = -\frac{1}{2000N}$ and
        \begin{equation*}
             w_{b_\gamma}^{(1,o_{\mathbf{t}_\gamma}^1)} = w_{b_\gamma}^{(1,o_{\mathbf{t}_\gamma}^1-1)} - \eta_{b_\gamma}
            \left. \frac{\partial\mathcal{L}}{\partial w_{c_\gamma}} \right|_{\mathbf{t}_\gamma} = -1 -2000N \times (-\frac{1}{2000N})=0
        \end{equation*}

        Note also that \texttt{clause($\gamma,\cdot,\cdot,\cdot$)}$\not\in T^1_{o_{\mathbf{t}_\gamma},|T|}$, by Lemma \ref{gd-weight-clause-missing} we have 
        \\$w_{c_\gamma}^{(1)}=w_{c_\gamma}^{(1,|T|)}=w_{c_\gamma}^{(1,o_{\mathbf{t}_\gamma}^1)}=\frac{1}{2}+\frac{1}{200N}$ and $w_{b_\gamma}^{(1)}=w_{b_\gamma}^{(1,|T|)}=w_{b_\gamma}^{(1,o_{\mathbf{t}_\gamma}^1)}=0$.

        (2)  If such $\mathbf{t}_\gamma=$\texttt{clause($\gamma,i_1,i_2,i_3$)} does not exist in $T$, by Lemma \ref{gd-weight-clause-missing} we have $w_{c_\gamma}^{(1)}=w_{c_\gamma}^{(0)}=\frac{1}{2}$ and $w_{b_\gamma}^{(1)}=w_{b_\gamma}^{(0)}=-1$.
    \end{proof}

    \begin{lemma}\label{gd-weight-epoch2-val-iter}
        Suppose $T\subseteq T_0$ and $C_l$ be the number of \texttt{clause}() in $T^2_{1,l}$. $\forall 1\le i\le n, 1\le l \le |T|$, $w_{x_i}^{(2,l)}\in U(1,\frac{C_l+1/2}{6N})$ if \texttt{var}($i$)$\in T$; Otherwise $w_{x_i}^{(2,l)}\in U(-1,\frac{C_l+1/2}{6N})$.
    \end{lemma}
    \begin{proof}
        Similar to the proof of \ref{gd-weight-epoch1-val-iter}, we prove this lemma by induction.
        
        \underline{Basic Case:} Note that for all $1\le i\le n$, $w_{x_i}^{(1)}=U(\pm 1,\frac{1}{6000N})$, and for all $1\le \gamma \le m, w_{c_\gamma}^{(1)}\in\{\frac{1}{2},\frac{1}{2}+\frac{1}{200N}\}, w_{b_\gamma}^{(1)}\in \{-1,0\}$.
        We denote $\mathbf{t}=\mathbf{t}^{(2,1)}$ to avoid cluttering. For any fixed $i$:

        (1) If $\mathbf{t}=$\texttt{var($i$)}, $C_1=0$. By Corollary \ref{gd-weight-epoch1-val}, $w_{x_i}^{(1)}=U(1,\frac{1}{6000N})$. We have 
        \begin{equation*}
            y_\mathbf{t} (\mathbf{w}^{(1)})^\top \mathbf{x}_\mathbf{t}^\prime = 5w_{x_i}^{(1)}\in U(5,\frac{1}{1200N})
        \end{equation*}
        hence $\left.\frac{\partial\mathcal{L}}{\partial w_{x_i}}\right|_\mathbf{t} = 0$, and 
        \begin{equation*}
            w_{x_i}^{(2,1)}=w_{x_i}^{(1)}\in U(1,\frac{1}{6N}) = U(1,\frac{C_l+1/2}{6N})
        \end{equation*}

        (2) If $\mathbf{t}=$\texttt{clause($\gamma,i,i^\prime,i^{\prime\prime}$)}, $C_1=1$. 
        By Lemma \ref{gd-weight-epoch1-clause}, we have $w_{c_\gamma}^{(1)}=\frac{1}{2}+\frac{1}{200N}$ and $ w_{b_\gamma}^{(1)}=0$.
        Therefore,
        \begin{equation*}
        \begin{aligned}
            y_\mathbf{t} (\mathbf{w}^{(1)})^\top \mathbf{x}_\mathbf{t}^\prime &= w_{x_i}^{(1)}+ w_{x_{i^\prime}}^{(1)}+ w_{x_{i^{\prime\prime}}}^{(1)} + w_{c_\gamma}^{(1)} + \frac{1}{2}w_{b_\gamma}^{(1)}\\
            &=  w_{x_i}^{(1)}+ w_{x_{i^\prime}}^{(1)}+ w_{x_{i^{\prime\prime}}}^{(1)} + \frac{1}{2} - \frac{1}{200N}\\
            &\in \bigcup_{x_0 \in \{\frac{1}{2}\pm 1,\frac{1}{2}\pm 3\}} U(x_0,0.01)
        \end{aligned}
        \end{equation*}
        hence $\left.\frac{\partial\mathcal{L}}{\partial w_{x_i}}\right|_\mathbf{t} \in \{0,-yx_{x_i} \} = \{-1,0\}$, and $\eta_{x_i}\left.\frac{\partial\mathcal{L}}{\partial w_{x_i}}\right|_\mathbf{t} \in \{-\frac{1}{6N},0\}$. 
        
        By Corollary \ref{gd-weight-epoch1-val}, if \texttt{var}($i$)$\in T$, we have
        \begin{equation*}
            w_{x_i}^{(2,1)}=w_{x_i}^{(1)} - \eta_{x_i}\left.\frac{\partial\mathcal{L}}{\partial w_{x_i}}\right|_\mathbf{t} \in U(1,\frac{3/2}{6N})=U(1,\frac{C_l+1/2}{6N})
        \end{equation*}
        
         If \texttt{var}($i$)$\not\in T$, we have
         \begin{equation*}
             w_{x_i}^{(2,1)}=w_{x_i}^{(1)} - \eta_{x_i}\left.\frac{\partial\mathcal{L}}{\partial w_{x_i}}\right|_\mathbf{t} \in U(-1,\frac{3/2}{6N})=U(-1,\frac{C_l+1/2}{6N})
         \end{equation*}
                 
        (3) Otherwise, $w_{x_i}$ will not be updated and $C_1 \le 1$. Therefore if  \texttt{var}($i$)$\in T$,
        \begin{equation*}
            w_{x_i}^{(2,1)}=w_{x_i}^{(1)}\in U(1,\frac{3/2}{6N})\subseteq U(1,\frac{C_l+1/2}{6N})
        \end{equation*}
        If \texttt{var}($i$)$\not\in T$,
        \begin{equation*}
            w_{x_i}^{(2,1)}=w_{x_i}^{(1)}\in U(-1,\frac{3/2}{6N})\subseteq U(-1,\frac{C_l+1/2}{6N})
        \end{equation*}

        Hence this lemma is true for $l=1$.

        \underline{Induction Step:} Suppose the lemma is true for $l<|T|$. We prove that this lemma remains true for $l+1$. We denote $\mathbf{t}=\mathbf{t}^{(2,l+1)}$ to avoid cluttering. This makes sense since $l+1\le|T|$ and thus $\mathbf{t}\in T$. For any fixed $i$:
        
        (1) If $\mathbf{t}=$\texttt{var($i$)}, $C_{l+1} = C_l$. By Corollary \ref{gd-weight-epoch1-val}, $w_{x_i}^{(2,l)}\in U(1,\frac{C_l+1/2}{6N})$.
        
        We have $y_\mathbf{t} (\mathbf{w}^{(2,l)})^\top \mathbf{x}_\mathbf{t}^\prime = 5w_{x_i}^{(2,l)}\in U(5,1/6)$
        and $\left.\frac{\partial\mathcal{L}}{\partial w_{x_i}}\right|_\mathbf{t} = 0$.Hence 
        $w_{x_i}^{(2,l+1)}= w_{x_i}^{(2,l)} \in U(1,\frac{C_{l+1}+1/2}{6N})$.
        
        (2) If $\mathbf{t}=$\texttt{clause($\gamma,i,i^\prime,i^{\prime\prime}$)}, $C_{l+1}=C_l + 1$. 
        In this case, \texttt{clause($\gamma,\cdot,\cdot,\cdot$)}$\not\in T_{1,l}^2$ and by Lemma \ref{gd-weight-clause-missing} and Lemma \ref{gd-weight-epoch1-clause} we have $w_{c_\gamma}^{(2,l)}=w_{c_\gamma}^{(1)}=\frac{1}{2}+\frac{1}{200N}, w_{b_\gamma}^{(2,l)}=w_{b_\gamma}^{(1)}=0$.    
        From the induction hypothesis we have $w_{x_i}^{(2,l)},w_{x_{i^\prime}}^{(2,l)},w_{x_{i^{\prime\prime}}}^{(2,l)}\in U(\pm 1, \frac{C_l+1/2}{6N})$. Noting that
        \begin{equation*}
        \frac{C_l+1/2}{6N}\le \frac{m+1/2}{6N}= \frac{m+1/2}{(n+2(m+1/2))} \le \frac{1}{12}
        \end{equation*}
        we have 
        \begin{equation*}
            \begin{aligned}
                y_\mathbf{t} (\mathbf{w}^{(2,l)})^\top \mathbf{x}_\mathbf{t}^\prime &= w_{x_i}^{(2,l)}+ w_{x_{i^\prime}}^{(2,l)}+ w_{x_{i^{\prime\prime}}}^{(2,l)} + w_{c_\gamma}^{(2,l)} + \frac{1}{2}w_{b_\gamma}^{(2,l)}\\
                &= w_{x_i}^{(2,l)}+ w_{x_{i^\prime}}^{(2,l)}+ w_{x_{i^{\prime\prime}}}^{(2,l)} +\frac{1}{2}+\frac{1}{200N}\\
                &\in \bigcup_{x_0 \in \{\frac{1}{2}\pm 1,\frac{1}{2}\pm 3 \}} U\left(x_0, \frac{3(C_l+1/2)}{6N}+\frac{1}{200N}\right)\\
                &\subseteq \bigcup_{x_0 \in \{\frac{1}{2}\pm 1,\frac{1}{2}\pm 3\}} U(x_0, 0.26)
            \end{aligned}
        \end{equation*}
        
        And thus $\left.\frac{\partial\mathcal{L}}{\partial w_{x_i}}\right|_\mathbf{t} \in \{0,-yx_{x_i} \} = \{-1,0\}$, and $\eta_{x_i}\left.\frac{\partial\mathcal{L}}{\partial w_{x_i}}\right|_\mathbf{t} \in \{-\frac{1}{6N},0\}$. 
        
        By Corollary \ref{gd-weight-epoch1-val}, if \texttt{var}($i$)$\in T$,
        $w_{x_i}^{(2,l+1)}=w_{x_i}^{(l)} - \eta_{x_i}\left.\frac{\partial\mathcal{L}}{\partial w_{x_i}}\right|_\mathbf{t} \in U(1,\frac{C_l+3/2}{6N})=U(1,\frac{C_{l+1}+1/2}{6N})$; if \texttt{var}($i$)$\not\in T$, $w_{x_i}^{(2,l+1)}=w_{x_i}^{(l)} - \eta_{x_i}\left.\frac{\partial\mathcal{L}}{\partial w_{x_i}}\right|_\mathbf{t} \in U(-1,\frac{C_l+3/2}{6N})=U(-1,\frac{C_{l+1}+1/2}{6N})$.
        
        (3) Otherwise, $w_{x_i}$ will not be updated. We have $C_{l+1}\le C_l + 1$ $w_{x_i}^{(2,l+1)}=w_{x_i}^{(2,l)}$. If \texttt{var}($i$)$\in T$ then $w_{x_i}^{(2,l+1)}\in U(1,\frac{C_{l+1}+1/2}{6N})$; If \texttt{var}($i$)$\not\in T$ then $w_{x_i}^{(2,l+1)}\in U(-1,\frac{C_{l+1}+1/2}{6N})$.

        Hence if the lemma is true for $l<|T|$, it is also true for $l+1$.
        Therefore, the lemma is true for all $1\le l\le |T|$.
    \end{proof}

    \begin{corollary}
    \label{gd-weight-epoch2-val}
    Suppose $T\subseteq T_0$ is the training data. 
    $\forall 1\le i\le n$, if \texttt{var($i$)}$\in T$, then $w_{x_i}^{(2)}\in U(1, 0.1)$. Otherwise $w_{x_i}^{(2)}\in U(-1, 0.1)$.
    \end{corollary}

    \begin{proof}
        Note that $w_{x_i}^{(2)} = w_{x_i}^{(2,|T|)}$ and $C_{|T|}\le m$. By Lemma \ref{gd-weight-epoch2-val-iter}, if \texttt{var($i$)}$\in T$ we have 
        \begin{equation*}
            w_{x_i}^{(2,|T|)}\in U(1,\frac{C_{|T|}+1/2}{6N}) \subseteq U(1,\frac{m+1/2}{6N})\subseteq U(1,\frac{1}{12}) \subseteq U(1,0.1)
        \end{equation*}
         If \texttt{var($i$)}$\not\in T$, we have 
         \begin{equation*}
            w_{x_i}^{(1,|T|)}\in U(-1,\frac{C_{|T|}+1/2}{6N}) \subseteq U(-1,\frac{m+1/2}{6N})\subseteq U(-1,\frac{1}{12}) \subseteq U(-1,0.1)
         \end{equation*}
    \end{proof}

    \begin{lemma}\label{gd-weight-epoch2-clause} Suppose $T\subseteq T_0$ is the training data. 
    $\forall 1\le i\le m$, if $\exists 1\le i_1,i_2,i_3 \le n$ such that \texttt{clause($i,i_1,i_2,i_3$)} $\in T$, then
        \begin{enumerate}
            \item $w_{b_j}^{(2)}=1000N$;
            
            \item $w_{c_j}^{(2)}=\frac{11}{2}+\frac{1}{200N}$ if exactly one of \texttt{var}($i_1$), \texttt{var}($i_2$), \texttt{var}($i_3$) is in $T$. Otherwise $w_{c_j}^{(2)}=\frac{1}{2}+\frac{1}{200N}$.
        \end{enumerate}
    Otherwise, $w_{b_i}^{(2)}=-1, w_{c_i}^{(2)}=\frac{1}{2}$.
    \end{lemma}

    \begin{proof}
        (1) If such $\mathbf{t}_\gamma=$\texttt{clause($\gamma,i_1,i_2,i_3$)} exists in $T$, by Lemma \ref{gd-weight-epoch2-val-iter} we have 
        \begin{equation*}
        w_{x_{i_1}}^{(2,o_{\mathbf{t}_\gamma}^1)},w_{x_{i_2}}^{(2,o_{\mathbf{t}_\gamma}^1)},w_{x_{i_3}}^{(2,o_{\mathbf{t}_\gamma}^1)}\in U(\pm 1,\frac{m+1/2}{6N})\subseteq U(\pm 1,\frac{1}{12N})
        \end{equation*}
        
        By Lemma \ref{gd-weight-clause-missing} we have $w_{c_\gamma}^{(2,o_{\mathbf{t}_\gamma}^1-1)}=w_{c_\gamma}^{(1)}=\frac{1}{2}+\frac{1}{200N}$ and $w_{b_\gamma}^{(2,o_{\mathbf{t}_\gamma}^1-1)}=w_{b_\gamma}^{(1)}=0$ because \texttt{clause($\gamma,\cdot,\cdot,\cdot$)}$\not\in T^1_{1,o_{\mathbf{t}_\gamma}-1}$. Consider the following two cases:
        
        (a) If exactly one of \texttt{var}($i_1$), \texttt{var}($i_2$), \texttt{var}($i_3$) is in $T$, by Corollary \ref{gd-weight-epoch2-val} we have
            \begin{equation*}
            \begin{aligned}
                 y_{\mathbf{t}_\gamma}(\mathbf{w}^{(2,o_{\mathbf{t}_\gamma}^1-1)})^\top\mathbf{x}_{\mathbf{t}_\gamma}^\prime &= 
                w_{x_{i_1}}^{(2,o_{\mathbf{t}_\gamma}^1-1)}+w_{x_{i_2}}^{(2,o_{\mathbf{t}_\gamma}^1-1)}+w_{x_{i_3}}^{(2,o_{\mathbf{t}_\gamma}^1-1)} + w_{c_\gamma}^{(2,o_{\mathbf{t}_\gamma}^1-1)} + \frac{1}{2}w_{b_\gamma}^{(2,o_{\mathbf{t}_\gamma}^1-1)} \\
                &=w_{x_{i_1}}^{(2,o_{\mathbf{t}_\gamma}^1-1)}+w_{x_{i_2}}^{(2,o_{\mathbf{t}_\gamma}^1-1)}+w_{x_{i_3}}^{(2,o_{\mathbf{t}_\gamma}^1-1)} + \frac{1}{2}+\frac{1}{200N} \\            
                &\in U(-\frac{1}{2},\frac{3}{12N}+\frac{1}{200N})\subseteq U(-\frac{1}{2},0.26)
            \end{aligned}
            \end{equation*}
        Hence $\left. \frac{\partial\mathcal{L}}{\partial w_{c_\gamma}} \right|_{\mathbf{t}_\gamma} = -1$, and
        \begin{equation*}
            w_{c_\gamma}^{(2,o_{\mathbf{t}_\gamma}^1)} = w_{c_\gamma}^{(2,o_{\mathbf{t}_\gamma}^1-1)} - \eta_{c_\gamma}
            \left. \frac{\partial\mathcal{L}}{\partial w_{c_\gamma}} \right|_{\mathbf{t}_\gamma} = \frac{1}{2}+\frac{1}{200N} +5=\frac{11}{2}+\frac{1}{200N}
        \end{equation*}

        Similarly, 
        \begin{equation*}
             w_{b_\gamma}^{(2,o_{\mathbf{t}_\gamma}^1)} = w_{b_\gamma}^{(2,o_{\mathbf{t}_\gamma}^1-1)} - \eta_{b_\gamma}
            \left. \frac{\partial\mathcal{L}}{\partial w_{b_\gamma}} \right|_{\mathbf{t}_\gamma} =1000N
        \end{equation*}

        Note also that \texttt{clause($\gamma,\cdot,\cdot,\cdot$)}$\not\in T^1_{o_{\mathbf{t}_\gamma},|T|}$, by Lemma \ref{gd-weight-clause-missing} we have 
        $w_{c_\gamma}^{(2)}=w_{c_\gamma}^{(2,|T|)}=w_{c_\gamma}^{(2,o_{\mathbf{t}_\gamma}^1)}=\frac{11}{2}-\frac{1}{200N}$ and $w_{b_\gamma}^{(2)}=w_{b_\gamma}^{(2,|T|)}=w_{b_\gamma}^{(2,o_{\mathbf{t}_\gamma}^1)}=1000N$.
        
        (b) Otherwise, we have
            \begin{equation*}
            \begin{aligned}
                 y_{\mathbf{t}_\gamma}(\mathbf{w}^{(2,o_{\mathbf{t}_\gamma}^1-1)})^\top\mathbf{x}_{\mathbf{t}_\gamma}^\prime &= 
                w_{x_{i_1}}^{(2,o_{\mathbf{t}_\gamma}^1-1)}+w_{x_{i_2}}^{(2,o_{\mathbf{t}_\gamma}^1-1)}+w_{x_{i_3}}^{(2,o_{\mathbf{t}_\gamma}^1-1)} + w_{c_\gamma}^{(2,o_{\mathbf{t}_\gamma}^1-1)} + \frac{1}{2}w_{b_\gamma}^{(2,o_{\mathbf{t}_\gamma}^1-1)} \\
                &=w_{x_{i_1}}^{(2,o_{\mathbf{t}_\gamma}^1-1)}+w_{x_{i_2}}^{(2,o_{\mathbf{t}_\gamma}^1-1)}+w_{x_{i_3}}^{(2,o_{\mathbf{t}_\gamma}^1-1)} + \frac{1}{2}+\frac{1}{200N} \\            
                &\in \bigcup_{x_0\in \{-\frac{7}{2},\frac{1}{2},\frac{5}{2}\}} U(x_0,\frac{3}{12N}+\frac{1}{200N})\subseteq \bigcup_{x_0\in \{-\frac{7}{2},\frac{1}{2},\frac{5}{2}\}} U(x_0,0.26)
            \end{aligned}
            \end{equation*}
        Hence $\left. \frac{\partial\mathcal{L}}{\partial w_{c_\gamma}} \right|_{\mathbf{t}_\gamma}= \left. \frac{\partial\mathcal{L}}{\partial w_{b_\gamma}} \right|_{\mathbf{t}_\gamma} = 0$, so $w_{c_\gamma}^{(2,o_{\mathbf{t}_\gamma}^1)} = w_{c_\gamma}^{(2,o_{\mathbf{t}_\gamma}^1-1)} = \frac{1}{2}+\frac{1}{200N}, w_{b_\gamma}^{(2,o_{\mathbf{t}_\gamma}^1)} = w_{b_\gamma}^{(2,o_{\mathbf{t}_\gamma}^1-1)} =0$.

        Note also that \texttt{clause($\gamma,\cdot,\cdot,\cdot$)}$\not\in T^1_{o_{\mathbf{t}_\gamma},|T|}$, by Lemma \ref{gd-weight-clause-missing} we have 
        $w_{c_\gamma}^{(2)}=w_{c_\gamma}^{(2,|T|)}=w_{c_\gamma}^{(2,o_{\mathbf{t}_\gamma}^1)}=\frac{1}{2}+\frac{1}{200N}$ and $w_{b_\gamma}^{(2)}=w_{b_\gamma}^{(2,|T|)}=w_{b_\gamma}^{(2,o_{\mathbf{t}_\gamma}^1)}=0$.

        (2)  If such $\mathbf{t}_\gamma=$\texttt{clause($\gamma,i_1,i_2,i_3$)} does not exist in $T$, by Lemma \ref{gd-weight-clause-missing} and  Lemma \ref{gd-weight-epoch1-clause} we have $w_{c_\gamma}^{(2)}=w_{c_\gamma}^{(1)}=\frac{1}{2}$ and $w_{b_\gamma}^{(2)}=w_{b_\gamma}^{(1)}=-1$.
    \end{proof}
    
    Moreover, $\mathbf{w}$ reaches its fixpoint at the end of the second epoch and will no longer be updated.

    \begin{lemma}\label{gd-weight-epoch-3}
    $\mathbf{w}^{(2)}=\mathbf{w}^{(3)}$.
    \end{lemma}
    
    \begin{proof}
        Suppose $\mathbf{w}^{(2)}\neq\mathbf{w}^{(3)}$, then there exists $1\le i\le N$ such that $w^{(2)}_i \neq w^{(3)}_i$, and there are some training sample $\mathbf{t}$ in the training data such that $\left.\frac{\partial\mathcal{L}}{\partial w_i^{(2)}}\right|_{\mathbf{t}}\neq 0$. Let $\mathbf{t}=(\mathbf{x_t},y_\mathbf{t})$ and $\mathbb{I}=U(-5,0.01)\cup  U(-\frac{1}{2},0.26)\cup \left(\bigcup_{x_0\in\{\pm 1,\pm 3\}} U(x_0,0.01)\right)$. By (\ref{gd-loss-derivative}) we have $y_\mathbf{t}(\mathbf{w}^{(2)})^\top\mathbf{x_t}^\prime\in \mathbb{I}$. 
        At least one of the following is true:

        \begin{enumerate}
            \item $\exists 1\le i\le n, \mathbf{t}=\texttt{var($i$)}$. According to lemma \ref{gd-weight-epoch2-val}, $y_\mathbf{t}(\mathbf{w}^{(2)})^\top\mathbf{x_t}^\prime = yw_{x_i}^{(2)}x_i \in U(5, 0.5)\subseteq \mathbb{R}\setminus\mathbb{I}$, contradicting to $y_\mathbf{t}(\mathbf{w}^{(2)})^\top\mathbf{x_t}^\prime\in \mathbb{I}$.

            \item $\exists 1\le i\le m$ and $1\le i_1,i_2,i_3\le n$, such that $\mathbf{t}=\texttt{clause($i,i_1,i_2,i_3$)}$.
            According to lemma \ref{gd-weight-epoch2-clause}, we have
            \begin{equation*}
                \begin{aligned}
                    y_\mathbf{t}(\mathbf{w}^{(2)})^\top \mathbf{x_t}^\prime &= w_{b_i}^{(2)}+w_{c_i}^{(2)}+w_{x_{i_1}}^{(2)} + w_{x_{i_2}}^{(2)}+w_{x_{i_3}}^{(2)}\\
                    &\ge 1000N + \frac{1}{2} + \frac{1}{200N} + 3\times(-1-0.1) \\
                    &\ge 1000-3.3 \ge 996
                \end{aligned}
            \end{equation*}
            We have $ y_\mathbf{t}(w^{(2)})^\top \mathbf{x_t}^\prime \not\in \mathbb{I}$, another contradiction.
        \end{enumerate}

        Therefore $\mathbf{w}^{(2)}=\mathbf{w}^{(3)}$,  $\mathbf{w}$ reaches its fixpoint at the end of the second epoch. In other words, $\mathbf{w}^* = \mathbf{w}^{(2)}$.
    \end{proof}
    
    We are now ready to give a rigorous proof of theorem \ref{GD-NPC}.
    

    \begin{proof}[Proof of theorem \ref{GD-NPC}]
        It only suffices to prove the correctness of the reduction in section \ref{gd-hardness}.
        
        \underline{If.} Suppose $\varphi\in$\textsc{Monotone 1-in-3 SAT}, then there is a truth assignment $\nu(\cdot)$ that assigns exactly one variable in each clause of $\varphi$ is true. Let $\Delta = \{\texttt{var}(i) | \nu(x_i)=\textsc{False} \}$. 
        Let $\mathbf{w}^\prime$ be the parameter of $\texttt{SGD}_\Lambda(T_0\setminus\Delta)$.
        By Lemma \ref{gd-weight-epoch2-clause}, $(w^\prime)_{c_\gamma}=\frac{11}{2}+\frac{1}{200N}$ for all $1\le \gamma \le m$, hence
        \begin{equation*}
            (\mathbf{w}^\prime)^\top\mathbf{x}_\text{test} = \sum_{\gamma=1}^m w_{c_\gamma}^\prime \ge \frac{11m}{2} + \frac{-11m+5}{2}=\frac{5}{2}>0
        \end{equation*}
        and $\lambda_{\mathbf{w}^\prime}(\mathbf{x}_\text{test})=1$, thus $\texttt{SGD}_\Lambda(T_0)$ is thus debuggable.  

        \underline{Only if.} Suppose $\texttt{SGD}_\Lambda(T_0)$ is debuggable, there will be a $\Delta$ such that $\texttt{SGD}_\Lambda(T_0,\mathbf{x}_\text{test})=y_\text{test}$ . We denote $\mathbf{w}^\prime$ as the parameter trained by \texttt{SGD} on $T_0\setminus\Delta$. We have $\lambda_{\mathbf{w}^\prime}(\mathbf{x}_\text{test})=1$ and $(\mathbf{w}^\prime)^\top \mathbf{x}_\text{test} \ge 0$.
        By Lemma \ref{gd-weight-epoch2-clause}, $w_{c_\gamma}^\prime = \{\frac{1}{2}+\frac{1}{200N},\frac{11}{2}+\frac{1}{200N}\}$. Suppose $w_{c^*}=\frac{1}{2}+\frac{1}{200N}$, then
        \begin{equation*}
        \begin{aligned}
             (\mathbf{w}^\prime)^\top\mathbf{x}_\text{test} &= w_{c^*} + \sum_{c_\gamma\neq c^*} w_{c_\gamma} \\
             &\le \frac{11}{2}(m-1)+\frac{1}{2}+\frac{m}{200N}-\frac{11m}{2}+\frac{5}{2} \\
             &= -\frac{5}{2} +\frac{m}{200N}\\
             &\le -\frac{5}{2}+\frac{1}{200}=-2.495<0
        \end{aligned}
        \end{equation*}
        leading to a contradiction.

        As a consequence, $w_{c_\gamma}^\prime = \frac{11}{2}+\frac{1}{200N}$ for all $1\le \gamma\le m$. By Lemma \ref{gd-weight-epoch2-clause}, exactly one of \texttt{var}($i_1$),\texttt{var}($i_2$),\texttt{var}($i_3$) is in $T_0\setminus\Delta$ for each $c_\gamma = (x_{i_1}\vee x_{i_2}\vee x_{i_3})$.
        Consider a truth assignment $\nu$ that maps every $x_i$ to \textsc{False} where \texttt{var}($i$)$\in \Delta$, and maps the rest to \textsc{True}. Then $\nu$ assigns exactly one variable true in each $c_\gamma = (x_{i_1}\vee x_{i_2}\vee x_{i_3})$ if and only if exactly one of \texttt{var}($i_1$),\texttt{var}($i_2$),\texttt{var}($i_3$) is in $T_0\setminus\Delta$. 
        Hence $\nu$ is a truth assignment that assigns true to exactly one variable in each clause of $\varphi$, and thus $\varphi$ is a yes-instance of $\textsc{Monotone 1-in-3 SAT}$. 
    \end{proof}

\section{Detailed Proofs for Section \ref{fixed-loss}}\label{fix-proof}
\subsection{Proof of Theorem \ref{1d-hinge-hard}}
\begin{proof}
    We build a reduction from the \textsc{Subset Sum} problem with a fixed size, which is NP-hard as a particular case of the class of knapsack problems \cite{subset}. Formally, it is defined as:

    \begin{center}
    \begin{minipage}{0.8\textwidth}
    \begin{mdframed}[
        linewidth=1pt,
        roundcorner=5pt
    ]
    \textsc{Subset Sum} with a fixed size
    
    \textbf{Input:} A set of positive integer $S$, and two positive integers $t,k$.

    \textbf{Output:} ``Yes'': if $\exists S^\prime\subseteq S$ of size $k$ such that $\sum_{a\in S^\prime}a = t$;\\
    \hspace*{3.7em}``No'': otherwise.
    \end{mdframed}
    \end{minipage}
    \end{center}

    The ordered training data $T$ is constructed as
    \begin{equation*}
        T=\{(x_1,y_1),(x_2,y_2),\dots,(x_n,y_n)\}\cup\{(x_a,y_a)\}
    \end{equation*}
    where $x_iy_i=\frac{2}{3}+\frac{a_i}{3\sum_{a\in S}a}$ for all $1\le i\le n$ and $x_ay_a=1+\frac{1}{6\sum_{a\in S}a}$. Let $\eta=1,\alpha=1,\beta=-1$, $w^{(0)}=-1-\frac{2}{3}k-\frac{t}{3\sum_{a\in S}a}$ and let the test instance $(x_\text{test},y_\text{test})$ satisfy $x_\text{test}y_\text{test}=1$. It now suffices to prove that $\exists S'\subseteq S$ such that $|S'|=k$ and $\sum_{a\in S'}a=t$ if and only if $\exists T'\subseteq T$ such that $w : w^{(0)} \xrightarrow{T^\prime}w$ satisfies $y_\text{test}wx_\text{test}>0$.
    
    \underline{If:} Suppose $\exists S'\subseteq S$ such that $|S'|=k$ and $\sum_{a\in S}a=t$. Let $T^*=\{(x_i,y_i)|a_i\in S'\}$, we prove that $y_\text{test}w^*x_\text{test}>0$ for $w^*$ satisfying $w^{(0)}\xrightarrow{T'=T^*\cup\{(x_a,y_a)\}}w^*$.
    
    Since
    $$
    \begin{aligned}
        w^{(0)}+\sum_{a_i\in S'}x_iy_i&=-1-\frac{2}{3}k-\frac{t}{3\sum_{a\in S}a}+\sum_{a_i\in S'}\left(\frac{2}{3}+\frac{a_i}{3\sum_{a\in S}a}\right)\\
        &=-1-\frac{2}{3}k-\frac{t}{3\sum_{a\in S}a}+\sum_{a_i\in S'}\frac{2}{3}+\frac{\sum_{a\in S'}a}{3\sum_{a\in S}a}=-1
    \end{aligned}
    $$
    and $\forall 1\le i \le n, x_iy_i>\frac{2}{3}$, for each $1\le i< n$, suppose $w^{(0)}\xrightarrow{T^*\cap\{(x_j,y_j)|1\le j\le i\}}w_i$, we have
    $$
    w_ix_{i+1}y_{i+1}< \left(w^{(0)}+\sum_{a_j\in S'}x_jy_j-\frac{2}{3}\right)\cdot\frac{2}{3}<-\frac{10}{9}<\beta.
    $$
    That is, each training sample in $T^*$ is activated. Then for $w^{(0)}\xrightarrow{T^*}w_a$, we have $w_a=-1$. Then, since $y_aw_ax_a=-(1+\frac{1}{6\sum_{a\in S}a})<\beta$ and $w_a\xrightarrow{(x_a,y_a)}w^*$ we have $w^*=w_a+x_ay_a=\frac{1}{6\sum_{a\in S}a}$. Therefore, $y_\text{test}w^*x_\text{test}=\frac{1}{6\sum_{a\in S}a}>0$.
    
    \underline{Only if:} For each $T'\subseteq T$, let $T^*=T'\setminus\{(x_a,y_a)\}$ and $c(T^*)$ be the set of training samples in $T^*$ that are activated. If $y_\text{test}w^*x_\text{test}\ge0$ for $w^*$ satisfying $w^{(0)}\xrightarrow{T'}w^*$, we prove that the set $S'=\{a_i|(x_i,y_i)\in c(T^*)\}$ satisfies $|S'|=k$ and $\sum_{a\in S'}a=t$.

    We first show that $y_\text{test}w_ax_\text{test}<0$ for $w^{(0)}\xrightarrow{c(T^*)}w_a$. Otherwise, suppose $y_\text{test}w_ax_\text{test}\ge 0$ we have $w_a\ge 0$. Let $(x,y)$ be the last training sample of $c(T')$, since $\frac{2}{3}< xy\le 1$, we have $w'\ge w_a-xy\ge-1$ for $w'\xrightarrow{(x,y)}w_a$. Thus $yw'x\ge\beta$, which contradicts to the definition of $c(T^*)$.

    We next show that $|S'|=k$. Suppose $|S'|\le k-1$, we have
    $$
    \begin{aligned}
        w_a=w^{(0)}+\sum_{(x_i,y_i)\in c(T^*)}x_iy_i&=-1-\frac{2}{3}k-\frac{t}{3\sum_{a\in S}a}+\sum_{a_i\in S'}\frac{2}{3}+\frac{\sum_{a\in S'}a}{3\sum_{a\in S}a}\\
        &<-1-\frac{2}{3}k+\frac{2}{3}(k-1)+\frac{1}{3}=-\frac{4}{3}
    \end{aligned}
    $$
    Thus $w^*\le w_a+x_ay_a<-\frac{4}{3}+(1+\frac{1}{6\sum_{a\in S}a})<0$ and then $y_\text{test}w^*x_\text{test}<0$, which contradicts to the fact that $y_\text{test}w^*x_\text{test}\ge 0$. Therefore $|S'|\ge k$.

    Suppose $|S'|\ge k+1$, we have
    $$
    \begin{aligned}
        w_a=w^{(0)}+\sum_{(x_i,y_i)\in c(T^*)}x_iy_i
        &\ge-1-\frac{2}{3}k-\frac{1}{3}+\frac{2}{3}(k+1)=-\frac{2}{3}
    \end{aligned}
    $$
    Then $y_aw_ax_a\ge (-\frac{2}{3})\cdot(1+\frac{1}{6\sum_{a\in S}a})\ge -\frac{7}{9}\ge \beta$, that is, $(x_a,y_a)$ is not activated and $w^*=w_a$. Then since $y_\text{test}w_ax_\text{test}< 0$, we have $y_\text{test}w^*x_\text{test}=y_\text{test}w_ax_\text{test}<0$, which contradicts to the fact that $y_\text{test}w^*x_\text{test}\ge 0$. Therefore $|S'|=k$.

    It remains to prove that $\sum_{a\in S'}a=t$. Otherwise, suppose $\sum_{a\in S'}a\le t-1$, we have
    $$
    \begin{aligned}
        w_a=w^{(0)}+\sum_{(x_i,y_i)\in c(T^*)}x_iy_i
        &\le-1-\frac{2}{3}k-\frac{t}{3\sum_{a\in S}a}+\frac{2}{3}k+\frac{t-1}{3\sum_{a\in S}a}\\
        &=-1-\frac{1}{3\sum_{a\in S}a}
    \end{aligned}
    $$
    Thus $y_\text{test}w^*x_\text{test}\le y_\text{test}(w_a+x_ay_a)x_\text{test}\le -\frac{1}{6\sum_{a\in S}a}<0$, which contradicts to the fact that $y_\text{test}w^*x_\text{test}\ge 0$. Therefore $\sum_{a\in S'}a\ge t$.

    Suppose $\sum_{a\in S'}a\ge t+1$ we have
    $$
    \begin{aligned}
        w_a=w^{(0)}+\sum_{(x_i,y_i)\in c(T^*)}x_iy_i
        &\ge-1-\frac{2}{3}k-\frac{t}{3\sum_{a\in S}a}+\frac{2}{3}k+\frac{t+1}{3\sum_{a\in S}a}\\
        &=-1+\frac{1}{3\sum_{a\in S}a}
    \end{aligned}
    $$
    Thus
    $$
    \begin{aligned}
        y_aw_ax_a&\ge (-1+\frac{1}{3\sum_{a\in S}a})\cdot(1+\frac{1}{6\sum_{a\in S}a})\\
        &\ge -1+\frac{1}{6\sum_{a\in S}a}+\frac{1}{18(\sum_{a\in S}a)^2}\ge \beta.
    \end{aligned}$$
    That is, $(x_a,y_a)$ is not activated and $w^*=w_a$. Then since $y_\text{test}w_ax_\text{test}<0$, we have $y_\text{test}w^*x_\text{test}=y_\text{test}w_ax_\text{test}<0$, which contradicts to the fact that $y_\text{test}w^*x_\text{test}\ge 0$. Therefore $\sum_{a\in S'}a=t$.
\end{proof}

\subsection{Proof of Theorem \ref{thm:hinge-hard} for \texorpdfstring{$\beta< -1$}.}
\begin{proof}

    To avoid cluttering, we still assume $\boldsymbol{\eta}=\mathbf{1}$ and $\alpha=1$. The proof can be generalized by appropriately re-scaling the constructed vectors.
  
    Let $M = -\beta(n+2) +9\beta nm^2(n+1) + 3$.
    Suppose $n=|S|>1$, $m=\max_{a\in S}\{a\}$ and $S=\{a_1,a_2,\dots,a_n\}$. We further assume $n>1$.
    Let the ordered set of training samples be
    \begin{equation*}
        T=\{(\mathbf{x}_1,y_1),(\mathbf{x}_2,y_2),\dots,(\mathbf{x}_n,y_n)\}\cup\{(\mathbf{x}_c,y_c),(\mathbf{x}_b,y_b),(\mathbf{x}_a,y_a)\}
    \end{equation*}
    where $\mathbf{x}_iy_i=(\frac{1}{n+1},-3\beta a_i) $ for all $1\le i\le n, \mathbf{x}_cy_c=(M+\frac{3}{2}\beta-1,\beta (3t -\frac{1}{2}) ),\mathbf{x}_by_b=(1,-1), \mathbf{x}_ay_a=(-\frac{3}{2}\beta,-\frac{3}{2}\beta)$.
    Let
    $\mathbf{w}^{(0)}=(-M,0)$.
    Let the test instance $(\mathbf{x}_\text{test},y_\text{test})$ satisfy $\mathbf{x}_\text{test}y_\text{test}=(1,0)$. 

    For each $1\le i<n$, suppose $\mathbf{w}^{(0)} \xrightarrow{T\cap \{(\mathbf{x}_i,y_i)|1\le j\le i\}}\mathbf{w}^{(i)}$, we have
      \begin{equation*}
          \begin{aligned}
              y_{i+1}(\mathbf{w}^{(i)})^\top\mathbf{x}_{i+1} &\le -M\cdot \frac{1}{n+1} + \frac{i}{(n+1)^2}+9\beta^2 a_{i+1}\sum_{j=1}^i a_j\\
              &\le -M\cdot \frac{1}{n+1}+\frac{n}{(n+1)^2} + 9\beta^2 nm^2 < \beta \\
          \end{aligned}
      \end{equation*}
      This means all the $(\mathbf{x}_i,y_i)\in T\setminus \{(\mathbf{x}_c,y_c),(\mathbf{x}_b,y_b), (\mathbf{x}_a,y_a)\}$ can be activated and thus the resulting parameter trained by $T\setminus \{(\mathbf{x}_c,y_c),(\mathbf{x}_b,y_b), (\mathbf{x}_a,y_a)\}$ is
      \begin{equation*}
          \mathbf{w}_c = \mathbf{w}^{(0)} + \sum_{i=1}^n\mathbf{x}_iy_i
          =\left( -M +\frac{|T^*|}{n+1} , -3\beta \sum_{i=1}^n a_i\right)
      \end{equation*}

    It now suffices to prove that for all $S^\prime\subseteq S$, $\sum_{a\in S^\prime} a = t$ if and only if $\exists T^\prime\subseteq T$ such that $\mathbf{w} : \mathbf{w}^{(0)} \xrightarrow{T^\prime}\mathbf{w}$ such that $y_\text{test}\mathbf{w}^\top\mathbf{x}_\text{test}>0$.

    \underline{If:} Suppose $\exists S^\prime\subseteq S$ such that $\sum_{a\in S}a=t$, we prove that $\exists T^\prime\subseteq T$ such that $y_\text{test}(\mathbf{w}^*)^\top\mathbf{x}_\text{test}>0$ for $\mathbf{w}^*$ satisfying $\mathbf{w}^{(0)}\xrightarrow{T^*}\mathbf{w}^*$.

    Let $T^*=\{(\mathbf{x}_i,y_i)|a_i\in S^\prime\}$, $T^\prime = T^*\cup 
    \{(\mathbf{x}_c,y_c),(\mathbf{x}_b,y_b), (\mathbf{x}_a,y_a)\}$. We have
    \begin{equation*}
        \mathbf{w}_c = (-M + \frac{|T^*|}{n+1},-3\beta \sum_{a_i\in S^\prime}a_i)= (-M + \frac{|T^*|}{n+1},-3\beta t)
    \end{equation*}
    And $y_c\mathbf{w}_c^\top\mathbf{x}_c=(-M+\frac{|T^*|}{n+1})(M+\frac{3}{2}\beta-1)-3t\beta^2(3t-\frac{1}{2})<\beta$, so
    \begin{equation*}
        \mathbf{w}_c \xrightarrow{(\mathbf{x}_c,y_c)}\mathbf{w}_b = \mathbf{w}_c + \mathbf{x}_cy_c = (\frac{|T^*|}{n+1}+\frac{3}{2}\beta-1,-\frac{1}{2}\beta)
    \end{equation*}
    Note that $\beta<-1$, we have $y_b\mathbf{w}_b^\top\mathbf{x}_b=\frac{|T^*|}{n+1}+2\beta< (\beta + \frac{|T^*|}{n+1}) + \beta<\beta$, and
    \begin{equation*}
        \mathbf{w}_b \xrightarrow{(\mathbf{x}_b,y_b)}\mathbf{w}_a = \mathbf{w}_b + \mathbf{x}_ay_a = (\frac{|T^*|}{n+1}+\frac{3}{2}\beta,-\frac{1}{2}\beta -1)
    \end{equation*}
    Note also that $y_a\mathbf{w}_a^\top\mathbf{x}_a=\frac{3}{2}(-\beta)(\frac{|T^*|}{n+1}-1+\beta)<\beta$, we have
    \begin{equation*}
        \mathbf{w}_a \xrightarrow{(\mathbf{x}_a,y_a)}\mathbf{w}^* = \mathbf{w}_a + \mathbf{x}_ay_a = (\frac{|T^*|}{n+1},-2\beta-1)
    \end{equation*}
    Therefore, $y_\text{test}(\mathbf{w}^*)^\top\mathbf{x}_\text{test}=\frac{|T^*|}{n+1}\ge 0$.

    \underline{Only if:} For each $T^\prime\subseteq T$, let $T^*=T'\setminus \{(\mathbf{x}_c,y_c),(\mathbf{x}_b,y_b), (\mathbf{x}_a,y_a)\}$, if $y_\text{test}(\mathbf{w}^*)^\top\mathbf{x}_\text{test}$ for $\mathbf{w}^*$ satisfying $\mathbf{w}^{(0)}\xrightarrow{T^\prime}\mathbf{w}^*$, we prove that $\exists S^\prime\subseteq S$ such that $\sum_{a\in S^\prime}a=t$. 
    We first show that for each $T^\prime\subseteq T$, if $\mathbf{w}(\mathbf{w}^{(0)}\xrightarrow{T^\prime}\mathbf{w})$ satisfying $y_\text{test}\mathbf{w}^\top\mathbf{x}_\text{test}\ge 0$, we have $\forall k\in\{a,b,c\},(\mathbf{x}_k,y_k)\in T^\prime,y_k\mathbf{w}_k^\top\mathbf{x}_k<\beta$, where $\mathbf{w}^{(0)}\xrightarrow{T^*}\mathbf{w}_c\xrightarrow{(\mathbf{x}_c,y_c)}\mathbf{w}_b\xrightarrow{(\mathbf{x}_b,y_b)}\mathbf{w}_a$. Otherwise, suppose $\exists k \in \{a,b,c\}$ such that $(\mathbf{x}_k,y_k)\not\in T^\prime$ or $y_k\mathbf{w}_k^\top\mathbf{x}_k\ge\beta$, we have
    \begin{equation*}
    \begin{aligned}
        y_\text{test}\mathbf{w}^\top\mathbf{x}_\text{test} &\le -M + \frac{|T^*|}{n+1}+M+\frac{3}{2}\beta -1 + 1 -\frac{3}{2}\beta - \min\left\{1,M+\frac{3}{2}\beta-1,-\frac{3}{2}\beta\right\}\\
        &=\frac{|T^*|}{n+1}-1<0
    \end{aligned}
    \end{equation*}
    which contradicts to the fact that $y_\text{test}\mathbf{w}^\top\mathbf{x}_\text{test}\ge 0$.

    Let $S^\prime=\{a_i|(\mathbf{x}_i,y_i)\in T^*\}$ and $t^\prime=\sum_{a\in S^\prime}a_i$, it suffices to prove $t^\prime=t$.
    Notice that 
    \begin{equation*}
    \begin{aligned}
        \mathbf{w}^{(0)}\xrightarrow{T^*}\mathbf{w}_c &= (-M+\frac{|T^*|}{n+1},-3\beta\sum_{a_i\in S^\prime}a_i)\\
        &=(-M+\frac{|T^*|}{n+1},-3\beta t^\prime)
    \end{aligned}
    \end{equation*}
    Hence $y_c\mathbf{w}_c^\top\mathbf{x}_c= (-M+\frac{|T^*|}{n+1})(M+\frac{3}{2}\beta-1)-3t^\prime\beta^2(3t-\frac{1}{2})<\beta$, thus
    \begin{equation*}
        \mathbf{w}_c\xrightarrow{(\mathbf{x}_c,y_c)}\mathbf{w}_b=\mathbf{w}_c + \mathbf{x}_cy_c = (\frac{|T^*|}{n+1}+\frac{3}{2}\beta-1,-3\beta(t^\prime-t)-\frac{1}{2}\beta)
    \end{equation*}

    (1) If $t^\prime\le t-1$, we have 
    \begin{equation*}
    \begin{aligned}
        y_b\mathbf{w}_b^\top\mathbf{x}_b &=\frac{|T^*|}{n+1}-1+2\beta+3\beta(t^\prime-t)\\
        &\ge \frac{|T^*|}{n+1}-(1+\beta)>0 >\beta
    \end{aligned}
    \end{equation*}
    a contradiction. Hence $\mathbf{w}_a = \mathbf{w}_b \xrightarrow{(\mathbf{x}_b,y_b)}\mathbf{w}_a = (\frac{|T^*|}{n+1}+\frac{3}{2}\beta,-3\beta(t^\prime-t)-\frac{1}{2}\beta-1)$.

    (2) If $t^\prime\ge t+1$, we have 
    \begin{equation*}
        \begin{aligned}
            y_a\mathbf{w}_a^\top\mathbf{x}_a&=-\frac{3\beta}{2}\left( \frac{|T^*|}{n+1}-1+\beta -3\beta(t^\prime-t)\right)\\
            &\ge -\frac{3\beta}{2}\left( \frac{|T^*|}{n+1}-1-2\beta \right)\\
            &>-\frac{3\beta}{2}\left( \frac{|T^*|}{n+1}+1 \right) >0>\beta
        \end{aligned}
    \end{equation*}
    another contradiction.
    Therefore $t^\prime = t$, and this completes the proof.

\end{proof}

\end{document}